# Advances and Outlooks of Heat Transfer Enhancement by Longitudinal Vortex Generators

Ya-Ling He[a] and Yuwen Zhang[a, b]

[a] Key Laboratory of Thermal-Fluid Science and Engineering of MOE，School of Energy and Power Engineering, Xi'an Jiaotong University, Xi'an, Shaanxi, 710049, P.R. China
[b] Department of Mechanical and Aerospace Engineering, University of Missouri, Columbia, MO 65211, USA

**Abstract**
In the last several decades, heat transfer enhancements using extended surface (fins) has received considerable attentions. A new heat transfer enhancement technique – longitudinal vortex generator (LVG) – received more and more attentions since the 1990s. It is a special type of extended surface that can generate vortices with axes parallel to the main flow direction. The vortices are generated as a result of strong swirling secondary flow caused by flow separation and friction. The state-of-the-art on the researches of LVG for its applications in heat transfer enhancement in straight channels, plate and wavy fin-and-tube heat exchangers, fin-and-oval-tube heat exchanger, and fin-and-tube heat exchangers with multiple rows of tubes. The trends and future directions on heat transfer enhancement LVG are discussed.

## 1. Introduction

Gas-gas and gas-liquid heat exchanger can find their wide applications in the areas of HVAC, refrigeration, electronics cooling, food processing, automobile, petroleum, aerospace and spaceflight and so on. The overall performances of the heat exchangers are often limited by low heat transfer coefficient on the gas side, which results in very low efficiency of energy utilization. The developments of modern industry and global energy shortage call for more compact heat exchangers with lower energy consumptions, as well lower vibration and noise. Thus, it is very imperative to develop heat transfer enhancement techniques with high-efficiency and low pressure drop to increase the heat transfer coefficient on the gas side.

Increase of the heat transfer coefficient on the gas side can be achieved by enhancing convective heat transfer on the gas side via various techniques: either passive or active [1]. The pass techniques, such as treated surfaces, rough surfaces, extended surfaces, swirl flow devices, and additives to the fluid, do not require any external powers. On the other hand, the active techniques, such as vibration, electromagnetic field and jet impingement, require external power to enhance heat transfer. In order to increase the heat transfer coefficient on the gas side, the cost-effective passive techniques, e.g., rough surfaces and extended surface, are being widely used. In the last several decades, heat transfer enhancement using extended surface (fins) has received considerable attentions. The traditional fins include wavy, perforated, slit, louvered, and composite fins, which alter the geometric configuration to improve the flow and heat transfer performances.

The heat transfer enhancement via various fins is accompanied by relatively significant increase of pressure drops. A new heat transfer enhancement technique – longitudinal vortex generator (LVG) – received more and more attentions since the 1990s. It is a special type of extended surface that can generate vortices with axes parallel to the main flow direction. The vortices are generated as a result of strong swirling secondary flow caused by flow separation and friction. In the traditional perspectives, passive heat transfer enhancement can be achieved by: (1) reducing boundary layer thickness, (2) swirling and flow destabilization, and (3) increasing the temperature gradient near the heat transfer surface. Longitudinal vortex generators can effectively take advantages of all three mechanisms for heat transfer enhancement. Compared to the traditional heat transfer enhancement techniques, the LVG can significantly increase the heat transfer coefficient on the gas side while the increase of pressure drop is mild. Due to the significant advantage of LVGs, researchers across the globe carried out systematic experimental and numerical studies on LVG and its applications in plate, channel, and fin-and-tube heat exchanges [2-17].

The state-of-the-art on the research of LVG for its applications in heat transfer enhancement will be reviewed first, followed by a detailed summary on the research works performed by our group. Finally, the trends and future directions on heat transfer enhancement LVG will be discussed.

## 2 Characteristics of Heat Transfer Enhancement by LVGs

To fulfill the requirements for environmentally friendliness, easy to use, conformability, and energy saving, the heat exchangers must have small size, low-noise, low power consumption, compact, and high stability. The heat exchangers must possess high heat transfer capacity so that under the given load, heat transfer can be accomplished under small temperature difference and lower flow velocity.

Under various constraints, the design goals can be met when the flow in the heat exchanger is close to laminar. From the perspective of the second law of thermodynamics, the entropy generations due to flow and heat transfer directly relates to the performance of the heat exchangers. Both large temperature difference and pressure drop can result in large entropy generation, which are not desirable. Therefore, the requirements of high heat transfer coefficient and low pressure drop coincide with the requirement of low entropy generation rate. From microscopic point of view, the rate of entropy generation in flow and heat transfer is related to the degree of disorder in the flow. Compared to the chaotic turbulent flow, the laminar flow is more ordered and therefore, using laminar flow in the heat exchanger is helpful to decrease the rate of entropy production and thus improve the energy efficiency of the heat exchangers. Traditional techniques, such as wave, slit, and louvered fins, enhance heat transfer by disturbing the flow, decreasing boundary layer thickness, or interrupting the development of boundary layer. These techniques can create small-sized transverse vortices whose axes are perpendicular to the main flow direction. As Fiebig [16] pointed out, the heat transfer enhancement by transverse vortices is limited for the case of steady-state laminar flow, while longitudinal vortices can significantly increase local and average heat transfer coefficients in the entire channel. Thus, LVGs have significant advantages for heat transfer enhancement in laminar flow in heat exchangers.

Figure 1 shows several common LVGs that include delta wings, rectangular wings, delta winglet, and rectangular winglet. As fluid passes LVGs, strong secondary swirling flow (see Fig 2) is generated and the tangential velocity of the vortices can be as high as two times of the main flow velocity. The high-velocity swirling secondary flow can not only promote mixing of the fluids in the main flow and edge regions, but can also "inject" the high-energy fluid into the boundary layer to suppress and delay the boundary layer separation, which decrease of profile drag. By special arrangement of LVGs, heat transfer in the air side of the heat exchanger is enhanced, while the pressure drop is decreased [8-9]. This appears to be counter-intuition but can be explained by the effect of LVGs on the profile drags on the fins and tubes. Although introduction of LVGs on the air side causes increase of profile drag for the fins, the specially arranged LVGs can delay flow separation on the tube so that the profile drag for the tube is decreased. The results reported in Ref. [8-9] could be obtained when decrease of the profile drag of the tubes is greater than the increase of profile drag of the fins.

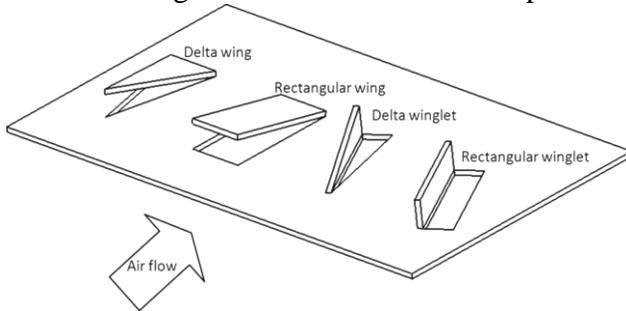 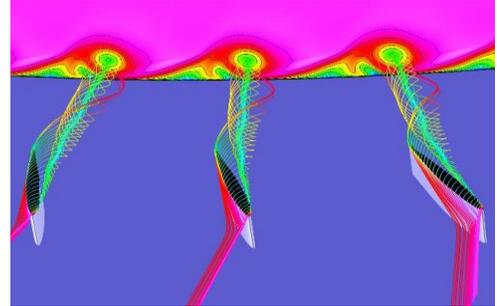

Fig. 1 Schematic diagrams of four common LVGs     Fig. 2 Generation of longitudinal vortices

## 3 Applications of LVGs on Heat Transfer Enhancement

The early studies on LVGs focused on their applications in the straight channel where the flow structures are relatively simple. While most early studies are done experimentally, more and more researchers numerically investigated applications LVGs in fin-and-tube heat exchangers, which were possible due to rapid development of computer hardware and software. The characteristics of flow and heat transfer can be thoroughly revealed by numerical simulation. Since there only two or three rows of tubes in the fin-and-tube heat exchangers in the residential air conditioning system, the number of rows in the early studies with LVGs are only two or three. Due to the continued growth of modern industry, the number of rows in the recent studies is mostly greater than three. The following discussions will focus on the applications LVGs in heat transfer enhancement in straight channels, plate and wavy fin-and-tube heat exchangers, fin-and-oval-tube heat exchanger, and fin-and-tube heat exchangers with multiple rows of tubes.

### 3.1 Heat Transfer Enhancement in Flat-Plate Channels by LVGs

Dupont et al. [18] experimentally studied isothermal flow in a compact heat exchanger channel with embossed-type vortex generators and concluded that the smooth shaped vortex generators are very promising for enhanced heat transfer. Gentry and Jacobi[19] carried out experiments on heat transfer in a channel with delta wings and found that the local heat transfer coefficient in the secondary flow region is increased by 300%; the average heat transfer coefficient in the channel is increased by 55% while the pressure drop is increased by 100%. Liou and Chen[20] investigated heat transfer and fluid flow in a square duct with 12 different shaped vortex generators using liquid crystal thermography and recommended LVGs with optimized geometric configurations. Wang et al. [21] experimentally studied heat transfer enhancement in narrow rectangular channel with longitudinal vortex generators using water as working fluid. The overal heat transfer capability was increased by 10 – 45% due to LVGs. Fiebig[16] experimentally studied flow and heat transfer characteristics in a rectangular channel with four types of LVGs and suggested that the longitudinal vortices resulted in a decrease of critical Reynolds number by one or more orders of magnitude. Compared to delta or rectangular wings, delta and rectangular winglets resulted in even lower pressure drops under the same heat transfer rate. Zhu et al. [22] numerically investigated turbulent flows and heat transfer in a rib-roughened channel with longitudinal vortex generators using k-ε model. The results showed that the combined effect of rib-roughness and vortex generators can enhance the average Nusselt number by nearly 450%.

Hiravennavar et al. [23] numerically solved the flow and heat transfer enhancement in a channel with built-in rectangular winglet pair and found that the average Nusselt number for the entire channel was increased by 66.6% due to the rectangular winglet pair. Biswas et al. [17] determined the flow structure and heat transfer effects of longitudinal vortices in a channel flow numerically and experimentally. The flow structure at the downstream of the LVGs were analyzed and the angle of attack with optimized heat transfer performance (*j/f*) was obtained.

Based on the existing research in the reported in the literatures, we systematically studied heat transfer enhancement by rectangular and delta winglets LVGs under common flow down (CFD) and common flow up (CFU) arrangements [24-25]. The mechanisms of heat transfer enhancement by LVGs were analyzed using the field synergy principle. The physical models for rectangular and delta winglets under two different arrangements were established (see Fig. 3) and the flow and heat transfer was numerically studied for isothermal wall and LVGs. A pair of LVGs are placed in the channel by either CFD or CFU arrangements. The working fluid was air with Prandtl number of 0.7. The shaded area in Fig. 3 was chosen as the computational domain due to symmetric condition.

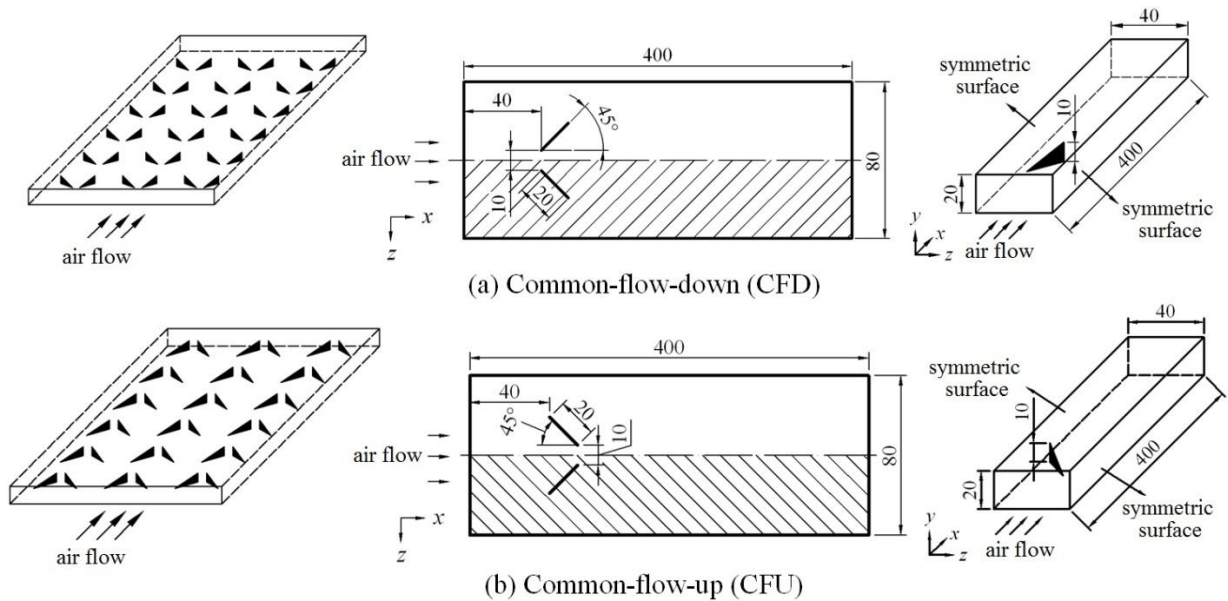

Fig. 3 Fluid flow and heat transfer in a channel with rectangular or delta winglets LVGs

Figure 4 shows the comparison of the ratio between average Nusselt number in a channel with LVGs (Nu) and smooth channel ($Nu_0$). It can be seen that the Nusselt number is increased by 4 ~ 16% after adding LVGs in the channel. Under the same aspect ratio, the heat transfer enhancement by rectangular winglets is more significant than that of the delta winglets. The Nusselt number for CFU arrangement is higher than that for CFD arrangement. Thus, the effect of arrangement on Nusselt number for rectangular winglet is more significant than that for the delta winglet.

Variations of the drag coefficient versus Reynolds number for different structures are shown in Fig. 5. It can be seen that the drag coefficient for the channel with LVGs are higher than that of the smooth channel. The drag coefficient for the channel with rectangular winglets is much larger than that that with delta winglets. For both rectangular and delta winglets, the drag coefficient for CFD arrangement is larger than that for the CFU arrangement. For delta winglets, the effect of arrangement on drag coefficient is not significant. On the contrary, the drag coefficient of the channel with rectangular winglets is more sensitive to how the winglets are arranged.

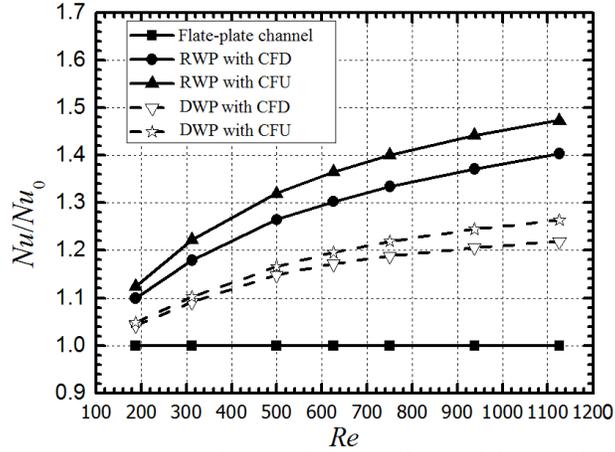
Fig. 4 Average Nusselt number vs. Reynolds number

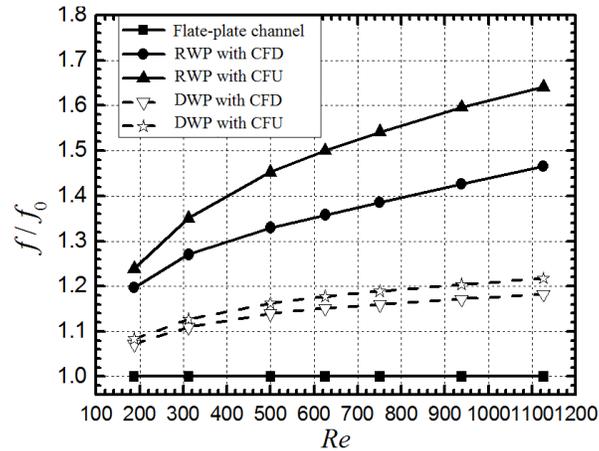
Fig. 5 Drag coefficient vs. Reynolds number

## 3.2 Heat Transfer Enhancement by LVGs in Fin-and-Tube Heat Exchangers

### 3.2.1 Fin-and-Tube Heat Exchanger with Two Rows of Staggered Tube Banks

Sommers and Jacobi[26] added delta winglet LVGs to the evaporator of the refrigerator and experimentally studied the heat transfer performance. The results showed that the thermal resistance on the air side is decreased by 35 to 42% when the Reynolds number is between 500 and 1300. Leu et al. [27] experimentally and numerically studied heat transfer and fluid flow in fin-and-tube heat exchangers with a pair of block shaped vortex generators. The results showed that the opyimized heat transfer performance is achieved when the atack amgle of the longitudinal vortices is 45°: the Colburn $j$ factor is increased by 8 – 30% while the drag coefficient, $f$, is only increased by 11 – 25%, which can achieve maximum saving of the heat transfer area by 25%. Allison and Dally [28] performed experimental studies the effect of a delta-winglet vortex generator pair on the performance of a tube-fin heat exchangers. Compared with the louvered fin-and-tube heat exchanger with the same size, the heat transfer capacity of the tube-and-fin heat exchanger is 87% of that of the louvered fin-and-tube heat exchanger, while the pressure drop the the former is only 53% of the latter. Zhang et al. [29] compared heat transfer performance of tube bank fin with mounted vortex generators to tube bank fin with punched vortex generators using naphetalane sublimation technique. The results showed that the that the effects of different LVGs on the heat transfer performance are very limited. Wu and Tao [30] investigated laminar convection heat transfer in fin-and-tube heat exchanger in aligned arrangement with LVG numerically. They also optimized the angle of attack of the LVG and conclued that the overall heat transfer performace is the best when the angle of attack is 30°.

The above literature review indicates that the application of LVGs on fin-and-tube heat exchanger with two rows of tubes is not thoroughly investigated although it is widely used in the modern residential air conditional system. Therefore, we studied hydrodynamics and heat transfer characteristics of a novel heat exchanger with two rows of tubes and delta-winglet vortex generators [31]. Figure 6 shows the schematic diagram of the computational domain of the fin-and-tube heat exchanger with two rows of tubes, while the sizes of the computational domain are shown in Fig. 7. The mechanisms of flow and heat transfer in the heat exchanger will be analyzed first, and the effects of tube arrangements, angle of attack of the LVG, and the aspect ratio of the delta winglet on the heat transfer performance will be studied.

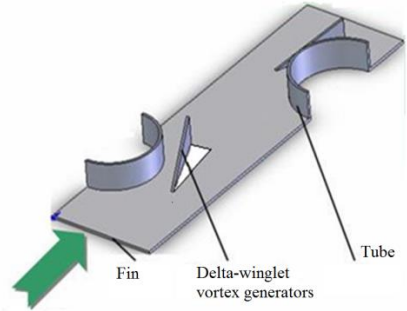

Fig. 6 Overview of the fin-and-tube heat exchanger and computational domain

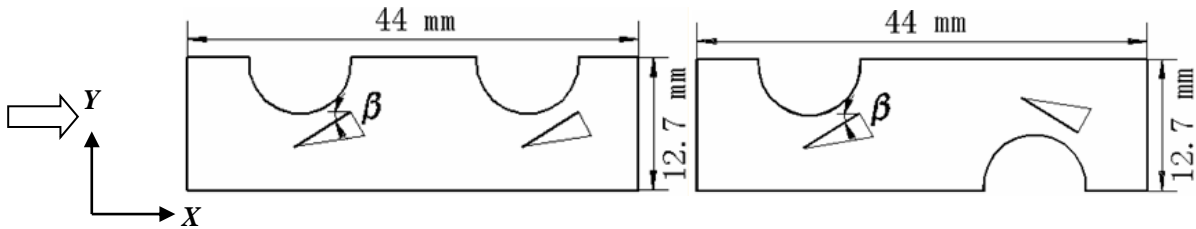

Fig. 7 Sizes of the computational domain

（1）**Effects of Side Delta Winglets on Flow and Heat Transfer**

The local flow field and temperature distribution in a fin-and-tube heat exchanger with side delta winglets will be analyzed first. It can be seen from Fig. 8 that a vortex whose axis is same as the main flow direction is generated. When the fluid flow across the delta winglet, the pressure difference before and after the delta winglet resulted in the secondary flow and generated the vortex. The vortex disturbed the fluid flow and reduced the boundary layer thickness. Figure 9 shows the velocity vector plot on a plane for both the enhanced and un-enhanced configurations at Re = 1000. The plane is close to the bottom fin surface at a normal distance of $z = 0.3$ mm from the bottom fin. It can be seen from Fig. 9 that the recirculation occurs widely behind the tubes, which further deteriorates the heat transfer. When the vortex generator with common-flow-up arrangement is employed, a nozzle-like passage appears between the tube and delta-winglet vortex generator. The fluid accelerates in the constrict passage and significantly delays the flow separation. Thus, the size of the wake zone and form-drag is significantly decreased. Thus, the addition of delta winglet not only generated longitudinal vortex, but also formed nozzle-like acceleration zone that decreased the size of the wake zone behind the tube.

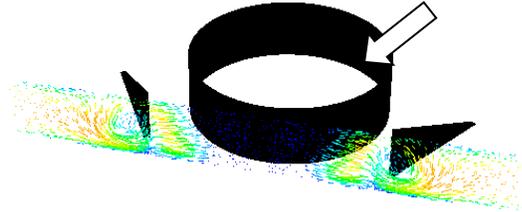

Fig. 8 Structure of longitudinal vortex ($Re$ = 1000)

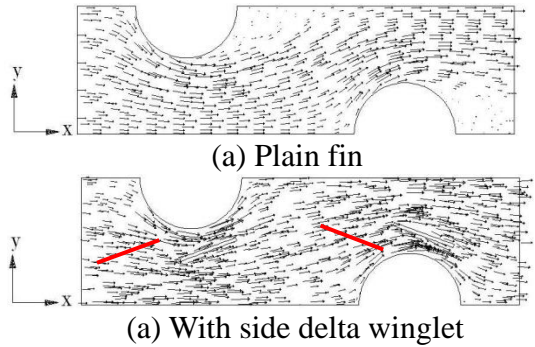

(a) Plain fin

(a) With side delta winglet

Fig. 9 Velocity vector plots at a plane near the fin (Re = 1000)

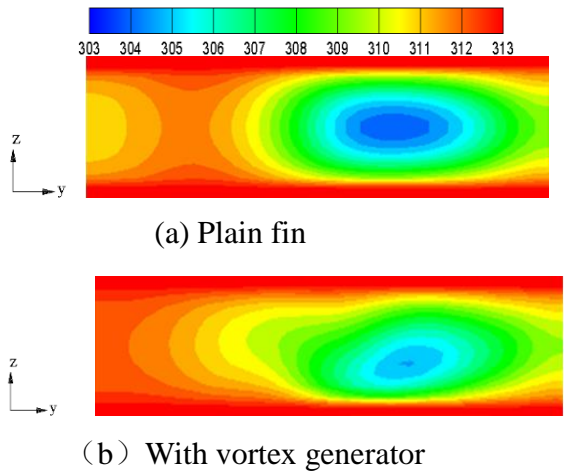

(a) Plain fin

(b) With vortex generator

Fig. 10 Exit temperature distribution (K)

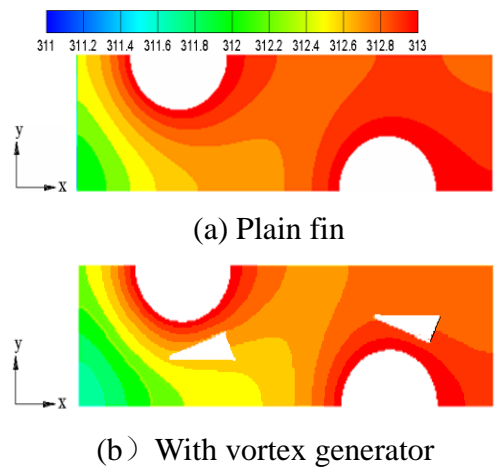

(a) Plain fin

(b) With vortex generator

Fig. 11 Fin surface temperature distribution (K)

Figure 10 shows the exit temperature distribution for plain fin and fin with delta winglets at $Re$ = 1000. It can be seen that the exit temperature for the case of plain fin is symmetric with respect to the middle-plane, but asymmetric for the case with vortex generator because the swirling flow rearranged the temperature distribution in the fluid. The air temperature gradient on the side with vortex generator is increased and the exit air temperature is higher than the case of plain fin. In other words, the difference between the air temperatures at inlet and outlet is increased. Consequently, the heat transfer rate with vortex generator is higher than the case of plain fin. Figure 11 shows the fin surface temperature distribution for plain fin and fin with delta winglet at Re = 1000. It can be seen that the temperature gradient in the region behind the tube for the case with vortex generator is higher than the case of plain fin. At the same location, the local temperature for the case with vortex generator is lower than that of the plain fin. The average fin temperature also lowers due to addition of vortex generator. Thus, the delta winglet increases the heat transfer rate and consequently increases the overall heat transfer coefficient so that the heat transfer performance is improved.

（**2**）**Effects of configuration and size of side delta winglet on the performance of fin-and-tube heat exchanger**

The effects of tube arrangement, angle of attack and aspect ratio of delta winglets on the heat transfer performance and pressure drop of will be analyzed below.

The tubes in the fin-and-tube heat exchangers can be arranged either inline or staggered. Figure 12 shows the variation of Colbum factor, $j$, as function of Reynolds number for different arrangements. At the same Reynolds number, the Colbum factor for the case with vortex generator is higher that that of the plain fin, regardless the arrangements. The mechanisms of heat transfer enhancement can be explained by the fact that the delta winglet generated the secondary flow and the fluids can directly impinge to the fin surface. The boundary layer becomes thinner and the hot and cold fluids can be well mixed. In addition, there exists an accelerated zone between the side delta winglet, tube wall, and fin, which led the fluid to enter the poor region for heat transfer behind the tube and reduced the size of the wake zone. In the range of Reynolds number that was studied, the heat transfer for inline arranged fin-and-tube heat transfer is enhanced by 38.8~50.9%, while the heat transfer enhancement for staggered arrangement is 35.1~45.2%. In other words, the heat transfer enhancement for inline arrangement is better than that for the staggered arrangement.

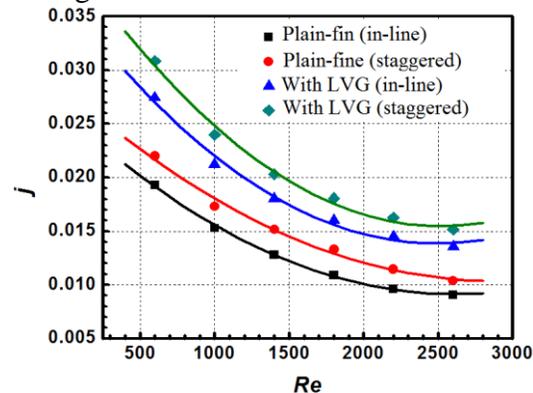 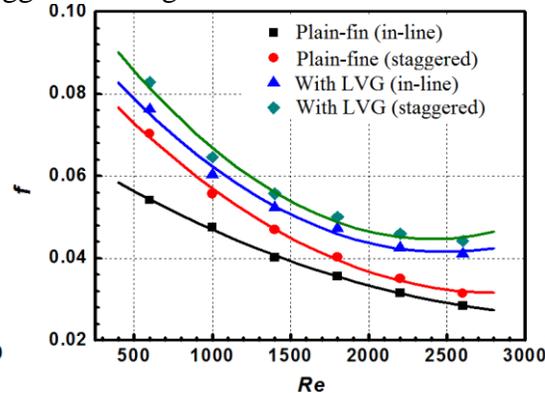

Fig. 12 Colburn factor vs. Reynolds number       Fig. 13 Drag coefficient vs. Reynolds number

Figure 13 shows the variation of drag coefficient with Reynolds number for different arrangements. It can bee seen that the drag coefficient decreases with increasing Reynolds number, and drag coefficient for the case with vortex generator is higher than that for the plain fin. For the fin-and-tube heat exchanger with side delta winglet, the change of the drag comes from two sources. The form drag of the delta winglet causes increase of the fin-and-tube heat exchanger. On the other hand, the delta winglet on the side of the tube resulted in increase of the fluid velocity near the tube and delayed the boundary layer separation; the size of the wake region is decreased so that the form drag of the tube is decreased. In the range of Reynolds numbers studied, the drag coefficient for inline arranged fin-and-tube heat transfer is increased by 30.3% ~ 46.8%, while the increase of drag coefficient for staggered arrangement is 19.3% ~ 34.5%. It can also be seen from Fig. 13 that increase of drag coefficient due to delta winglet at higher Reynolds number is more significant than that at the low Reynolds number.

The ratio of Colburn factor and drag coefficient, $j/f$, for different tube arrangement is shown in Fig. 14. For both inline and staggered arrangements, the overall performances for the cases with delta winglets are better than that of the plain fins, which indicate that the enhancement of heat transfer is more significant than increase of the drag coefficient. It should be noted that as Reynolds number increases, increase of $j/f$ due to delta winglet becomes less significant since increase of drag coefficient at higher Reynolds number is more significant than the case of low Reynolds number. Therefore, it can be concluded that the delta winglet is more effective on enhancing heat transfer at low Reynolds number.

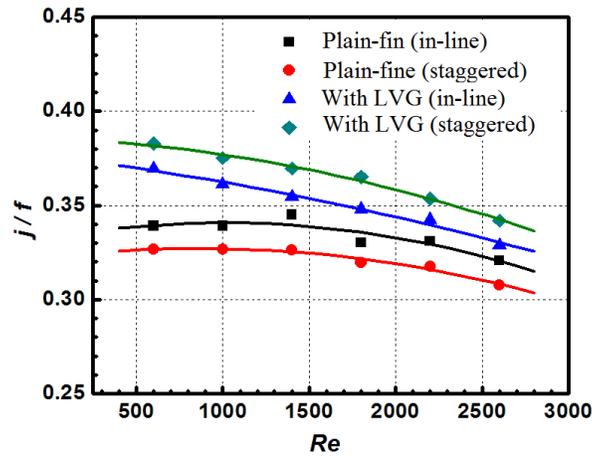

Fig. 14 Overall heat transfer and drag for different arrangements

In order to reveal the effects of angle of attack of the delta winglets on heat transfer and pressure drop, performance of fin-and-tube heat exchangers with delta winglets at five different angles of attack ($\beta = 10°、20°、30°、40°、$ and $50°$) are studied. Figure 15 shows the Coburn factor versus Reynolds at five different angles of attack. It can be seen that heat transfer is enhanced for all five cases. At the same Reynolds number, the Colburn factor increases with increasing angle of attack. Increase of Coburn factor with angle of attack is more significant at small angle of attack, and such increase becomes less significant at large angle of attack. Figure 16 shows drag coefficient versus Reynolds number at different angles of attack. It can be seen that drag coefficient is increased for all five cases and the drag coefficient increases with increasing angle of attack. Increase of drag coefficient with angle of attack is less significant at small angle of attack. As angle of attack further increases, the increase of drag coefficient becomes more significant.

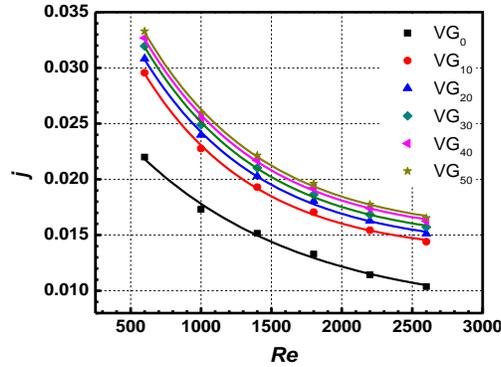

Fig. 15 Colburn factor vs. Reynolds number at different angles of attack

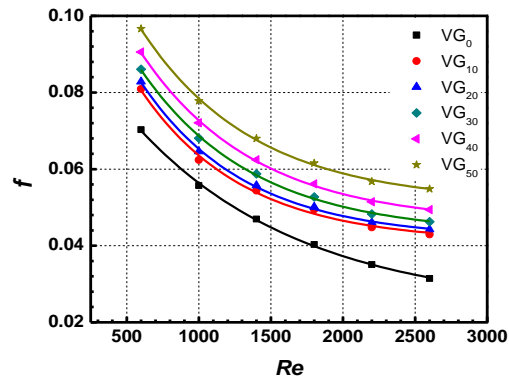

Fig. 16 Drag coefficient vs. Reynolds number at different angles of attack

The overall heat transfer and drag for delta winglet at different angle of attack are shown in Fig. 17. For the case that the angle of attack is 50°, the value of *j/f* is greater than the case of plain fin when the Reynolds number is under 1800. When the Reynolds number is above 1800, the *j/f* value for the case of delta winglet is lower than that of plain fin. For all other four angles of attack, the overall performances of the cases with delta winglet are better than that of the plain fin for all Reynolds numbers studied. It should be noted that that the increase of *j/f* becomes less significant as Reynolds number increases. Therefore, side delta winglet type LVG is more effective at low Reynolds number. The computational results also indicate that that, among all structures studied, the delta winglet with angle of attack of 20° has the best overall performance.

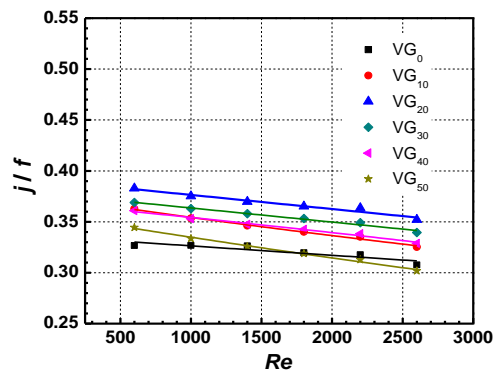

Fig. 17 Overall performances of heat transfer and drag at different angles of attack

The effects of aspect ratio of the delta winglet are studied while the angle of attack is fixed at 20°. Figure 18 shows the Colburn factor versus Reynolds at different aspect ratios, $\Lambda$. Under fixed Reynolds number, the Colburn factor increases with increasing aspect ratio, and the increase becomes less significant at higher aspect ratio. Figure 19 shows the drag coefficient versus Reynolds number at different aspect ratios. The drag coefficient increases with increasing aspect ratio and the increase becomes more pronounced at higher aspect ratio. The overall performances of heat exchanger with delta winglets at different aspect ratios are shown in Fig. 20. For all four aspect ratios that were studied, the value of $j/f$ decreases with increasing Reynolds number. The delta winglets with aspect ratio of 2 have the best overall performance on heat transfer and pressure drop.

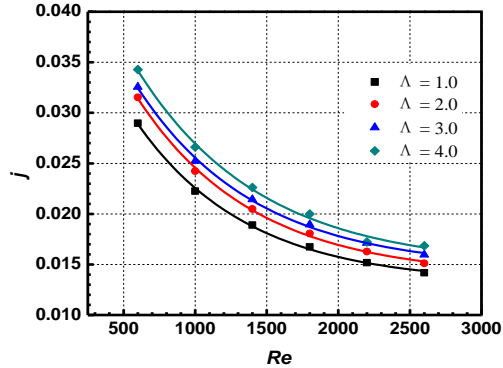

Fig. 18 Colburn factor vs. Reynolds number at different aspect ratio

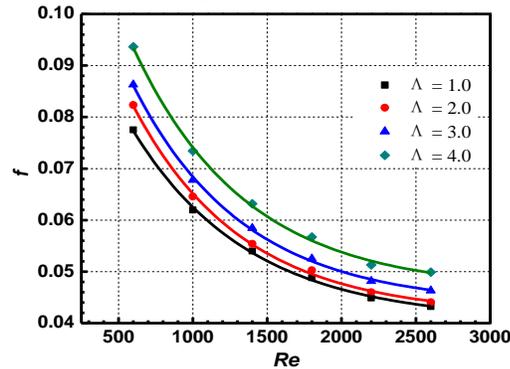

Fig. 19 Drag coefficient vs. Reynolds number at different aspect ratio

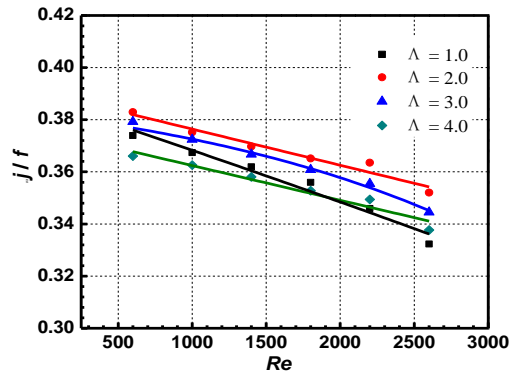

Fig. 20 Overall performances of heat transfer and drag at different aspect ratio

## 3.2.2 Application of LVGs in Wavy Fin-and-Tube Heat Exchangers

Combining LVGs with other heat transfer enhancement techniques can further improve the air-side heat transfer of the fin-and-tube heat exchangers. Sanders and Thole[32] experimentally studied the effects of winglets to augment tube wall heat transfer in louvered fin heat exchangers. The effects of the LVG's angle of attack, aspect ratio, direction, and shape on the performance of the fin-and-tube heat exchanger were investigated for 230 < Re < 1016. At optimized winglet parameters, tube wall heat transfer augmentations as high as 39% were achieved with associated drag coefficient augmentations as high as 23%. Combination of LVGs with wavy fin-and-tube heat exchangers has received scant attentions in the past and was studied by our group [33-34]. The flow and heat transfer characteristics of wavy fin-and-tube heat exchangers as shown in Fig. 21 was investigated. Effects of angle of attack, number of rows, and wavy angle on the performance of the wavy fin-and tube heat exchangers were discussed.

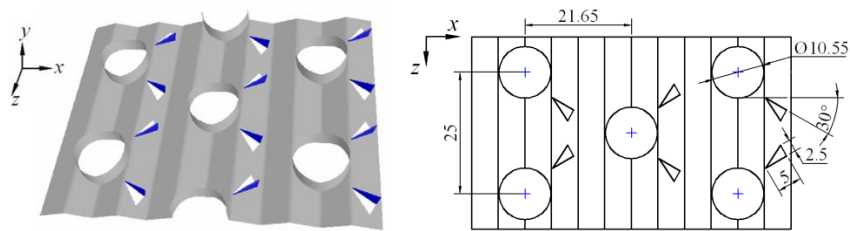

Fig. 21 Schematic of wavy fin with delta winglet

（1）**Effects of tube arrangements on flow and heat transfer**

Figure 22 shows the variation of the local average pressure along the streamwise direction. The local average pressure at any cross section is determined by the area-weighted average static pressure at this cross-section. There exists a steep pressure drop near the tube. For the wavy fin with delta winglets, the local average pressure has a slight drop at the axial location of the delta winglet, which is due to a small form drag induced by the slender delta winglet. The increase in the pressure drop penalty induced by the delta winglet is relatively small as indicated by Fig. 22.

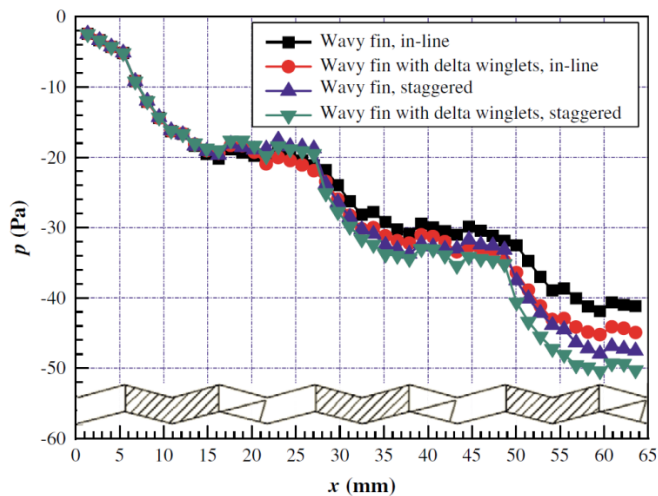

Fig. 22 Local average pressure along the streamwise direction

Figure 23 shows effects of longitudinal vortices generated by delta winglets on the local spanwise averaged local heat transfer coefficient distribution along the main flow direction. At the leading edges of fins of both inline and staggered arrangements, the heat transfer coefficient is the highest. However, the heat transfer weakens rapidly as boundary layer thickens. For the case of inline arrangement without delta winglet, there exists a large wake zones between tube rows where the heat flux is low. After delta winglets are added, the longitudinal vortices generated by delta winglets significantly enhanced heat transfer in the zone between tube rows, which is evidenced by a peak of heat transfer coefficient between the tube rows. For the case of staggered tube arrangement without delta winglets, it is different from the inline arrangement that fluid directly impinging to the tubes and the heat flux reaches to peak value at the stagnant point of every tube and then decreases as the boundary layer develops. Similar to the case of inline arrangement, there also exists a large wake zone behind every tube where heat transfer worsens. For the case with delta winglets, longitudinal vortices are generated due to separation of fluid; the wake zone becomes smaller and the local heat transfer is significantly enhanced. It is evidence from Fig. 23(b) that a new heat flux peak appeared behind the tube due to longitudinal vortices. Therefore, for both inline and staggered arrangement, the longitudinal vortices generated by delta winglets significantly enhanced heat transfer in the wake zone which compensate the heat transfer in this poorest region in the fin-and-tube heat exchangers.

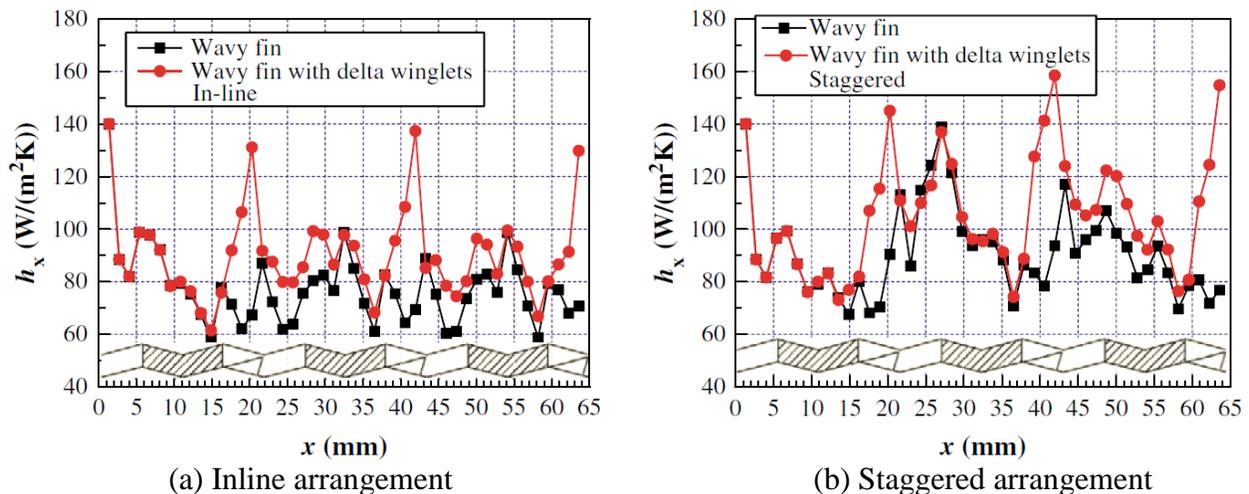

(a) Inline arrangement　　　　　　　　　　　(b) Staggered arrangement

Fig. 23 Distribution of spanwise averaged local heat transfer coefficient along the streamwise direction

(2) **Effect of angle of attack of LVG on flow and heat transfer**

Effects of angle of attack of the delta winglet on flow and heat transfer are investigated and compared to the cases with the same wavy fin-and-tube heat exchanger without LVGs. Figure 24 shows the variations of Colburn factor and drag coefficient with Reynolds number at different angles of attack. It can be seen that both $j$ and $f$ for the case with delta winglets are greater than the case without delta winglets. Meanwhile, both $j$ and $f$ increase with increasing angle of attack $\beta$. In the range of Reynolds numbers studied, the Colburn factor for the case with delta winglets and angle of attack of $\beta=30°$ is 8~12% higher than the case without delta winglets, while the fraction factor is only increased by 2~7%. At an angle of attack of $\beta=45°$, the Colburn factor and drag coefficient are respectively increased by 13~17% and 9~12% after addition of delta winglets. When the angle of attack is further increased to $\beta=60°$, the delta winglets resulted in increases of Colburn factor and drag coefficient by 17~21% and 19~21%, respectively. As the angle of attack increases, the projection area of the delta winglets normal to the incoming flow increases so that the form drag also increases. Meanwhile, the intensities of the longitudinal vortices also increases and their disturbances on flow are also stronger. As a result, the pressure drop increases with increasing angle of attack. Therefore, heat transfer enhances with increasing angle of attack, while pressure drop also increases with increasing angle of attack.

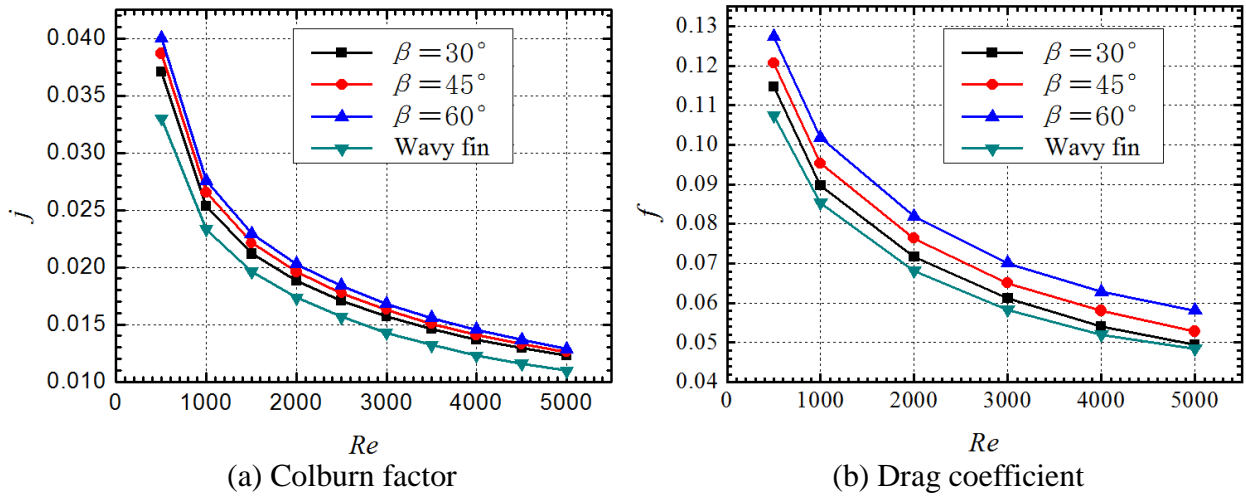

(a) Colburn factor  (b) Drag coefficient
Fig. 24 Effects of angle of attack on Colburn factor and drag coefficient

Figure 25 shows the overall performance of the wavy fin-and-tube heat exchanger versus angle of attack $\beta$. At the angle of attack of 30° and 45°, the ratio of Colburn factor and drag coefficient for the case with delta winglets are higher than the case without LVG; $j/f$ is the largest at 30°. At angle of attack of 60°, on the other hand, $j/f$ for the case with delta winglets is lower than that of the case without LVG for most Reynolds number except Re = 500. Therefore, at smaller angle of attack, the enhancement of heat transfer by LVGs is higher than the increase of the drag coefficient. As angle of attack increases, the cost of pressure drop outweighs the benefit of heat transfer enhancement.

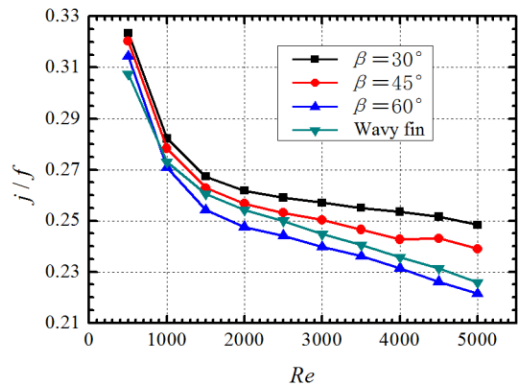

Fig. 25 Effect of angle of attack on $j/f$

（3）**Effect of number of rows on flow and heat transfer**

Effects of number of rows on flow and heat transfer in a wavy fin-and-tube heat exchanger with delta winglets are investigated. Figure 26 shows the variations of Colburn factor and drag coefficient with Reynolds number for different numbers of rows. The Colburn factor of the wavy fin-and-tube heat exchanger with delta winglets and four rows of tubes is slightly larger than that with two and three rows of tubes; the difference between the cases of two and three rows of tubes is very insignificant. The drag coefficient of the wavy fin-and-tube heat exchanger with delta winglets and two rows of tubes is slightly larger than that with three and four rows of tubes. Therefore, increasing number of rows results in slight increases of $j$ and decrease of $f$. One can conclude that due to combined wavy and delta winglets, the disturbance in the channel increases which causes flow and heat transfer becomes fully-developed. Consequently, the effect of the number of rows on flow and heat transfer is not very significant. Figure 27 shows the effect of number of row of tubes on $j/f$. It can be seen that, under the same Reynolds number, $j/f$ increases with increasing number of rows.

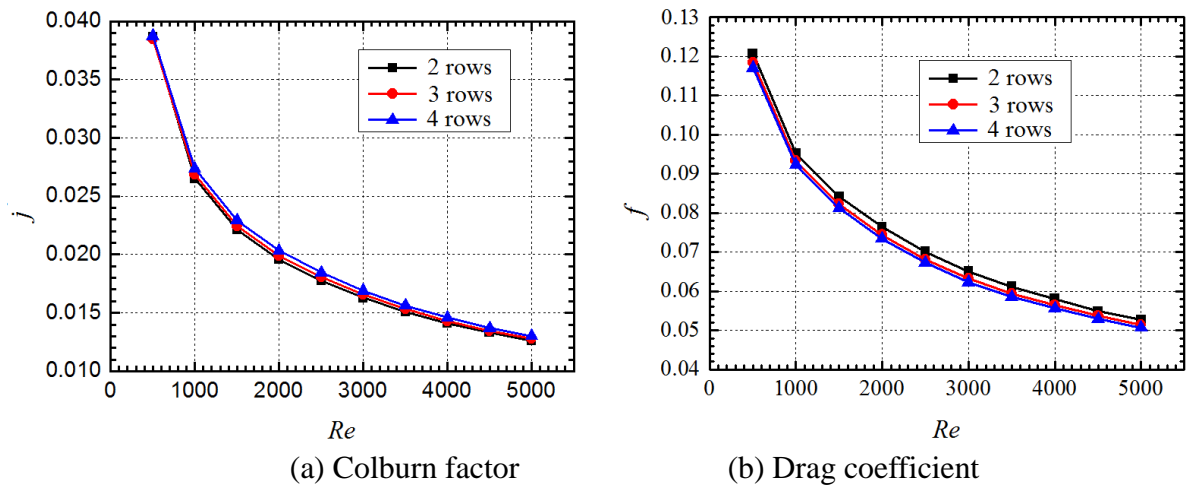

(a) Colburn factor  (b) Drag coefficient

Fig. 26 Effects of the number of rows on Colburn factor and drag coefficient

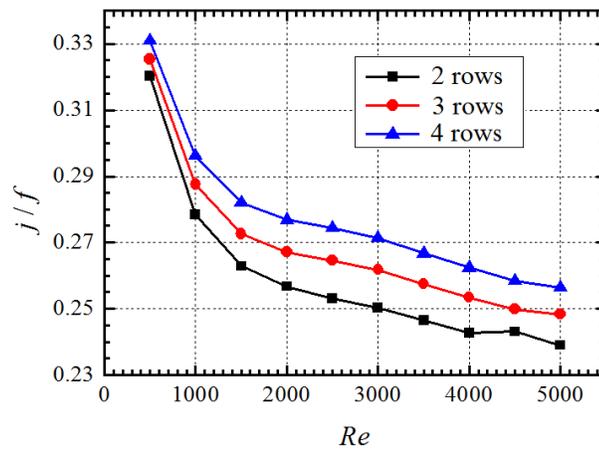

Fig. 27 Effect of number of rows on the overall performance

（4）**Effect of wavy angle on flow and heat transfer**

The flow and heat transfer at different wavy angles ($\theta = 0°$, 5°, 10°, 15°, and 20°) are studied. It should be pointed out that when $\theta = 0°$, the wavy fin with LVGs becomes plain fin with LVGs. Figure 28 shows variation of $j$ and $f$ with Reynolds number at different wavy angles. As wavy angle increases, the Colburn factor decreases until $\theta = 5°$ and then increases afterwards, i.e., the heat transfer at $\theta = 5°$ is the poorest. When the wavy angle is increased to $\theta = 10°$, the Colburn factor is slightly higher than the case of plain fin. The drag coefficient that reflects the flow characteristics increases with increasing wavy angle and the rate of increase becomes higher at larger wavy angle.

Figure 29 shows the effect of wavy angle on the overall performance of the heat exchanger. It can be seen that $j/f$ decreases with increasing wavy angle. In other words, the increase of pressure drop due to increasing wavy angle is more significant than enhancement of heat transfer. Therefore, the performance of the fin degrades as wavy angle increases.

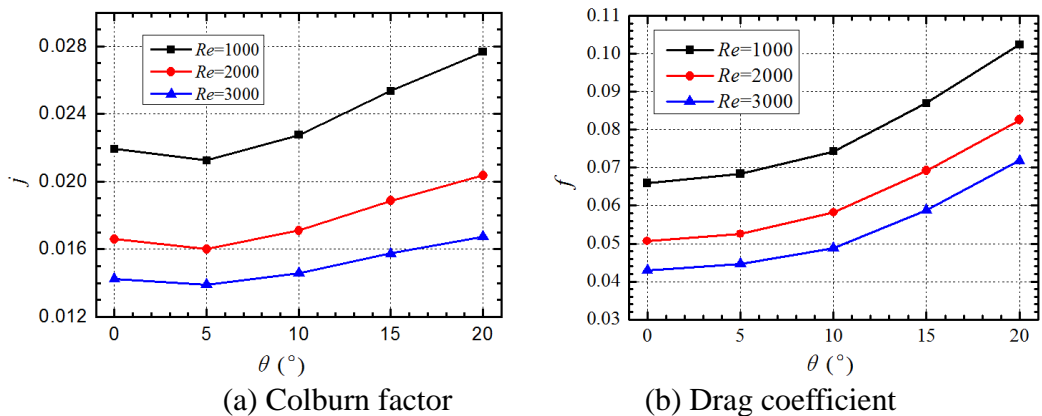

(a) Colburn factor  (b) Drag coefficient
Fig. 28 Effects of wavy angle on Colburn factor and drag coefficient

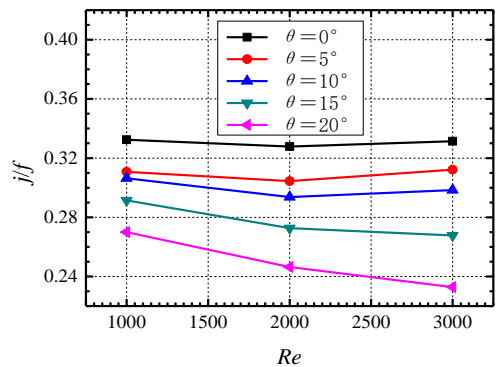

Fig. 29 Effect of wavy angle on the overall performance

### 3.2.3 Applications of LVGs and oval tubes on fin-and-tube heat exchangers

Since the oval tube has better performance on drag reduction and can effectively decrease the size of the wake zone behind the tube, the pressure drop can be effectively decreased. On the other hand, LVGs increases the form drag in the channel and results in increases of drag and pressure drop. Therefore, combination of oval tube and LVGs can take advantages of both of them. The heat transfer capacity can be substantially increased without too much pressure loss.

Chen and Fiebig[35] numerically simulated fin-and-oval-tube heat exchangers unit with delta winglets. The angle of attack and aspect ratio of the delta winglet were optimized at Re = 300. Their results showed that the optimized overall heat transfer performance [$(j/j_0)/(f/f_0)=1.04$] can be obtained when the angle of attack was 30° and the aspect ration was 2. Based on the above work, they thoroughly investigated effect of multiple LVGs on the performance of heat exchangers with inline[36] and staggered[37] arrangements of tubes. Tiwari et al. [38] numerically studied performance of fin-and-oval-tube heat exchanger unit with delta winglet and analyzed the flow and heat transfer under different numbers of LVGs (1~4 pairs) and arrangements (inline and staggered). Their results showed that the when two pairs of delta winglets are arranged inline, the average Nusselt number is increased by 43.86%; when four pairs of delta winglets are staggered, the average Nusselt number is increased by about 100%. O'Brien and Sohal [39] experimentally investigated flow and heat transfer in a narrow rectangular duct fitted with a circular tube and/or a delta-winglet pair. Their results showed that when a pair of delta winglets is installed, the average Nusselt of the rectangular duct with fitted oval tube was increased by 38%; while the corresponding drag was increased by 10% ($Re_h$ =500) and 5% ($Re_h$ =5000), respectively. Herpe et al. [40] umerically investigated the local entropy production rate of a finned oval tube with vortex generators.

The existing researches often focus on the performance of one heat transfer unit (e.g., one oval tube), while the detailed studies on the flow and heat transfer of the entire channel are lacking. We investigated flow and heat transfer in fin-and-tube heat exchangers with delta winglets and the effects of key parameters are studied [41-43].

(1) **Effects of LVGs on flow and heat transfer in fin-and-oval-tube heat exchangers**

In order to reveal the effects of LVGs on the overall flow and heat transfer performance of fin-and-tube heat exchangers, numerical simulations on fin-and-oval-tube heat exchangers with and without LVGs are performed. Figure 30 shows the schematic diagram of the fin-and-oval-tube heat exchangers with delta winglets. The LVGs are symmetrically installed behind the oval tube and the shaded area is the computational domain.

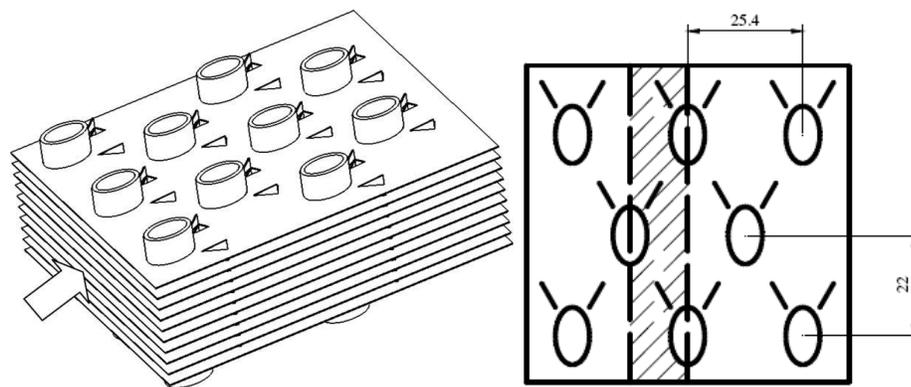

Fig. 30 Fin-and-oval tube heat exchangers with LVGs and the computational domain (unit: mm)

The flow channels of the fin-and-oval-tube heat exchangers without and with delta winglets are shown in Fig. 31. The locations and orientation of the LVGs are shown in Fig. 32.

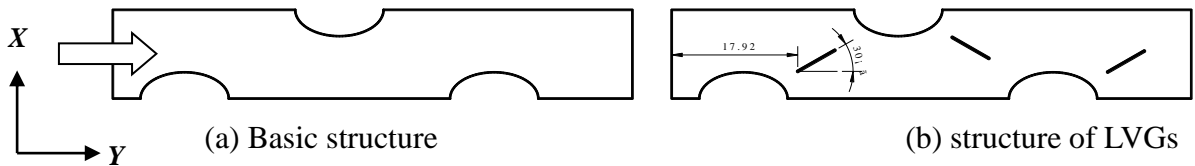

(a) Basic structure            (b) structure of LVGs

Fig. 31 Flow channel of the fin-and-oval-tube heat exchangers

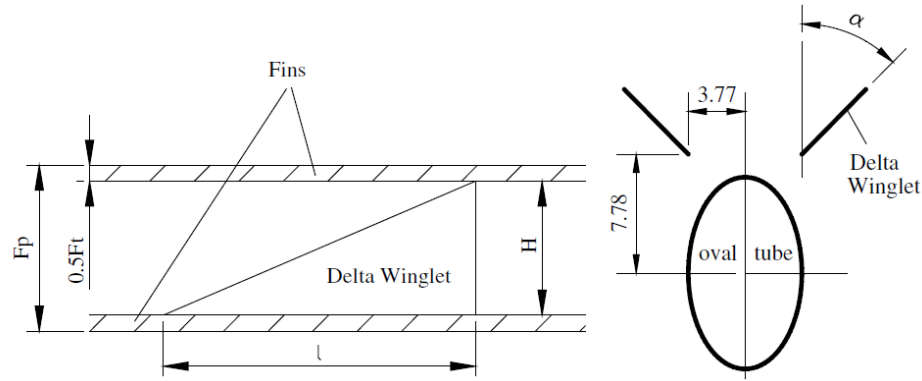

Fig. 32 Size and locations of the LVGs

    When air flows through the channel of the fin-and-oval-tube heat exchanger with LVGs, longitudinal vortices are generated due to the pressure difference before and after LVGs and the friction. The axis of this strong swirling secondary flow is same as the main flow direction. Due to strong disturbance of the LVGs, the boundary layers can be weakened or its formation can be interrupted. The strong funnel effects of the longitudinal vortices can also bring the fluid from the wake region to the main flow region. The cold fluid near the edge and the hot fluid in the main flow region can be well mixed and the heat transfer can be enhanced.

    Figure 33 shows the isovel distributions in three x–z planes at Re = 1500. The velocity in the entrance region before LVGs is nearly uniform and without any vortices. The variation of the velocity is also not very significant. After the fluid passes LVGs, the generation of longitudinal vortices resulted in highly non-uniform isovels and produced strong secondary flow. The transverse velocity can be as high as three times of the inlet velocity. The strong swirling flow transports the fluid near fin and tube wall to the core of main flow. Meanwhile, the fluid in the core of the main flow is also carried over to the region near fin and tube wall. These processes significantly promoted mixing of hot and cold fluids and increase the heat transfer coefficient.

    Figure 34 shows the velocity vector plots and streamlines at three cross-sections normal to the main flow direction. When the fluid passes LVGs, the pressure variation and the separation of the fluid at the LVG surface generate very complex swirling flow. As can be seen from Fig. 34, in addition to the main vortex, induced vortex and corner vortex can also be formed. The combined effects from various vortices resulted in complete disturbance of the thermal boundary layer. The hot and clod fluids are fully mixed and heat transfer is enhanced.

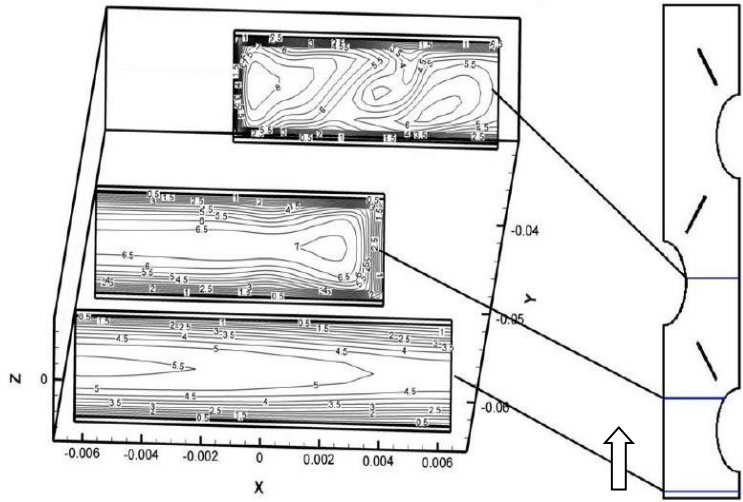

Fig. 33 Distributions of isovels in three cross-sections normal to the flow direction (unit: m/s)

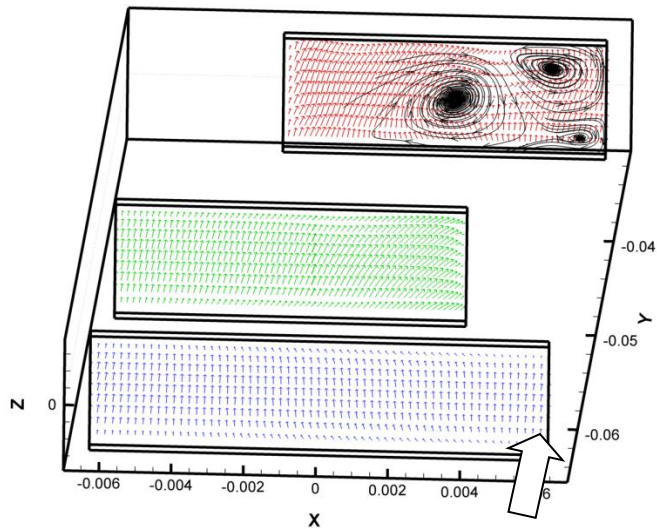

Fig. 34 Vector-plots and streamlines generated by LVGs in three cross-sections normal to the main flow direction

Figure 35 shows the temperature contour at three cross-sections normal to the main flow direction at Re =1500. In the entrance region, the isotherms are parallel to each other, and there is no apparent change on the thermal boundary layer before the fluid passing the LVGs. However, the isotherms are twisted and distorted after the LVGs. The thermal boundary layer becomes thinner and temperature gradient increases on the fin surface impinging by the longitudinal vortices. There changes causes increasing heat transfer coefficient on the fin surface and the heat transfer performance of the heat exchanger is improved.

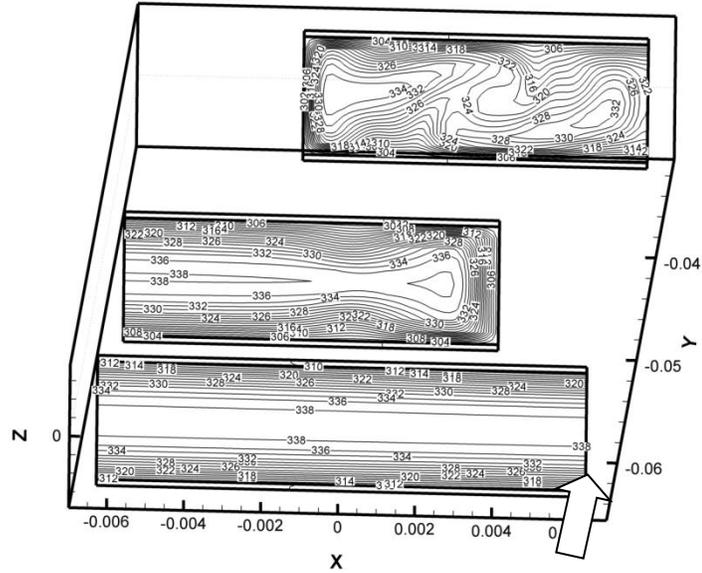

Fig. 35 Isotherms on three cross-sections of the normal to the main flow direction (unit: K)

Figure 36 shows the local velocity distribution on the middle cross-section (parallel to the x–y plane) for the cases without and with LVGs. It can be seen from Fig. 36(a) that there exists a large wake zone for the case without LVGs. The fluid in this zone is almost isolated from the fluid in the main flow. A thermal barrier is formed and heat transfer in this zone is extremely poor. After the LVGs are installed, the strong transverse secondary flow generated from the longitudinal vortices effectively reduced the size of the wake zone. Meanwhile, the fluid with high momentum is redirected to the oval tube surface by the longitudinal vortices, which, in turn, effectively delays the separation of boundary layer on the oval tube (see Fig. 36(b)). All of the above mechanisms can effectively contribute to the enhancement of heat transfer.

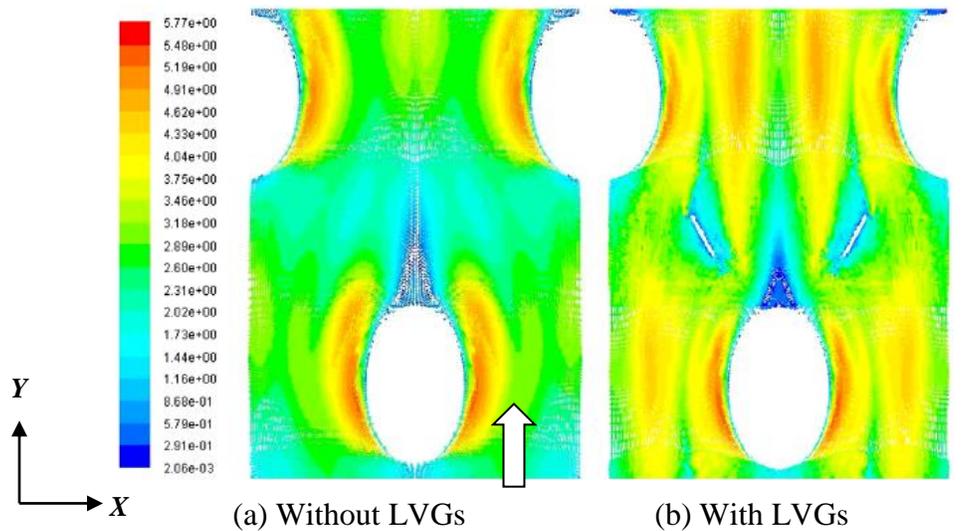

(a) Without LVGs  (b) With LVGs
Fig. 36 Local velocity distribution on the middle cross-section (unit: m/s)

Figure 37 shows the local temperature profiles on the middle cross-section for Re = 1500. It can be seen from Fig. 37(a) that the temperature in the aforementioned thermal barrier zone is close to that of the oval tube. The thermal barrier zone becomes significantly smaller after LVGs are installed (see Fig. 37(b)). Comparison of Fig. 37(a) and (b) indicates the temperature distributions before LVGs are almost the same for both cases. However, the fluid temperature is significantly lowered after the fluid passing LVGs, especially in the downstream region of the LVGs. The generation of the longitudinal vortices altered the flow field and promoted the mixing between the cold and hot fluids. The temperature gradient on the heat transfer surface is also increased, which ultimately resulted in heat transfer enhancement in the entire heat exchanger.

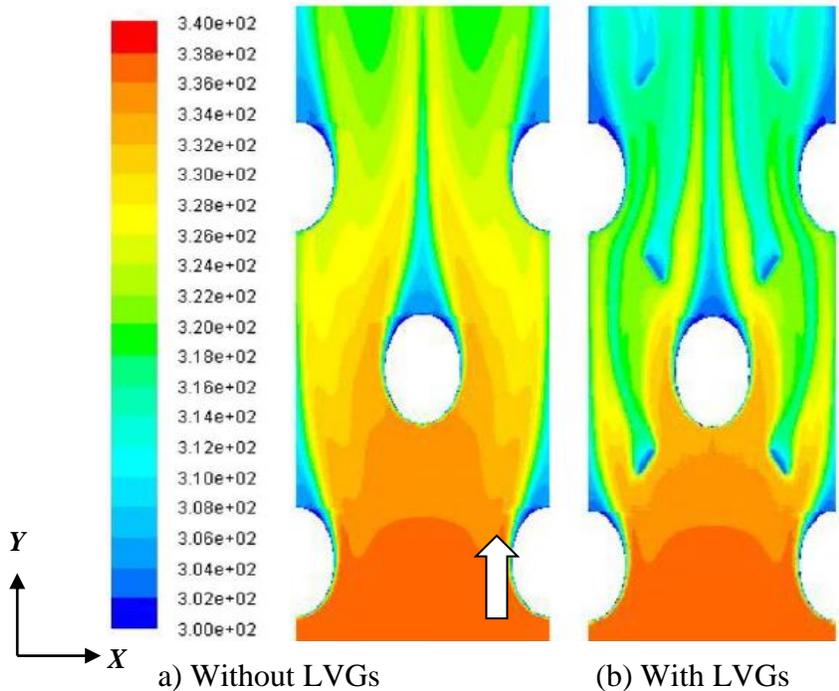

a) Without LVGs        (b) With LVGs
Fig. 37 Local temperature profiles on the middle cross-section (unit: K)

Figure 38 shows the average Nusselt number versus Reynolds number for the case without and with LVGs. It can be seen that both Nusselt numbers increase with increasing Reynolds number. In the range of Reynolds numbers studied (Re = 500 ~ 2500), the fin-and-oval-tube heat exchanger with LVGs showed better heat transfer performance over the case without LVGs. Compared to the case without LVGs, the average Nusselt number for the case with LVGs are 13.6 ~ 32.9 % higher. Figure 39 shows the drag coefficient versus Reynolds number for the case without and with LVGs. Both drag coefficients decrease with increasing Reynolds number. In the range of Reynolds numbers studied (Re = 500 ~ 2500), the fin-and-oval-tube heat exchanger with LVGs exhibited higher drag coefficient over the case without LVGs. Compared with the case without LVGs, the drag coeffiecent for the case with LVGs is 29.6 ~ 40.6 % higher. The reason for the increased drag confident is that the existences of LVGs increased the form drag so that the pressure drop for the heat exchanger is increased.

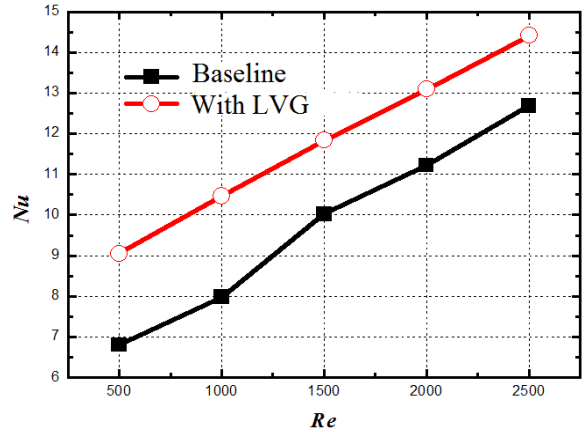

Fig. 38 Average Nusselt number versus Reynolds number

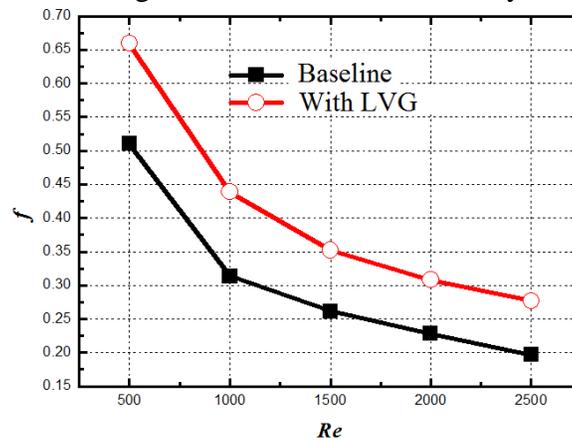

Fig. 39 Drag coefficient versus Reynolds number

    The simulation results are also analyzed by using the field synergy principle, where the intersection angle between velocity and temperature gradient is an important parameter. Figure 40 shows the average interaction angle versus Reynolds number. It can be seen that the average intersection angles for both cases decrease with increasing Reynolds number. This means that as Reynolds number increases, the disturbance becomes stronger and the angle between velocity vector and temperature gradient decreases. In other words, the synergy between the velocity and temperature fields is improved. In the range of Reynolds numbers studied (Re = 500 ~ 2500), the intersection angle for the fin-and-oval-tube heat exchanger with LVGs is always less than that for the case without LVGs. This means that the existence of LVGs improved the synergy between velocity filed and temperature in the heat exchanger and decreased the intersection angle, which led to enhancement of heat transfer performance.

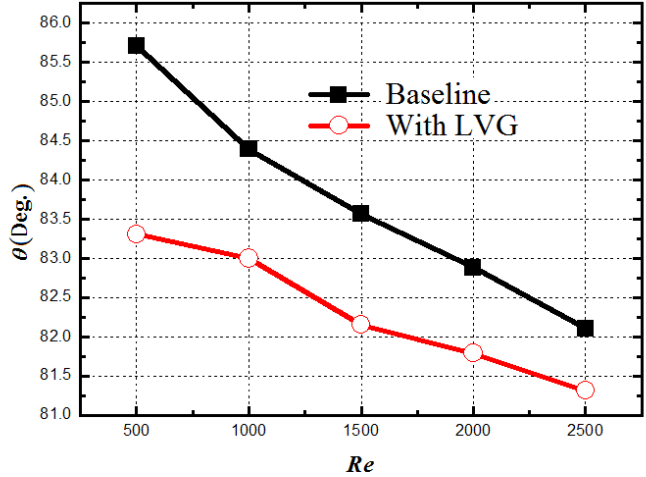

Fig. 40 Comparison of intersection angle between velocity vector and temperature gradient

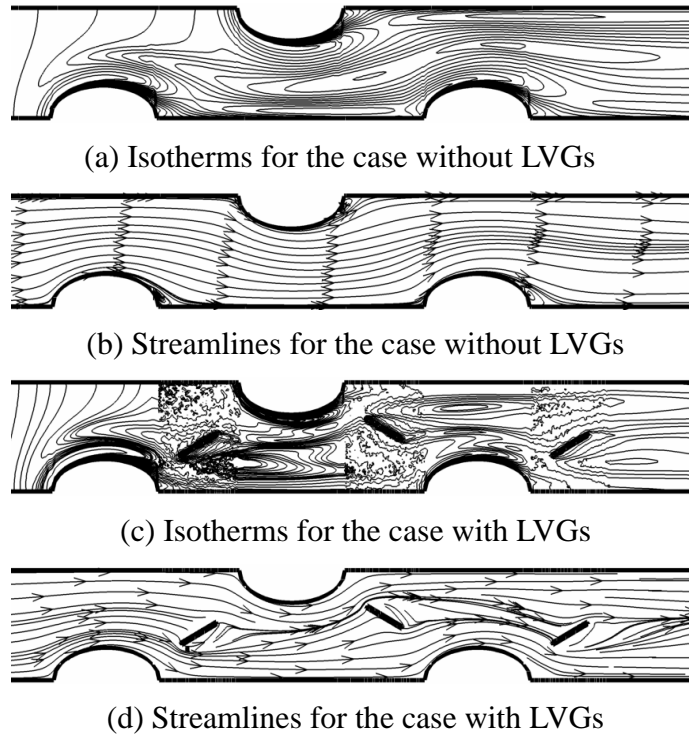

(a) Isotherms for the case without LVGs

(b) Streamlines for the case without LVGs

(c) Isotherms for the case with LVGs

(d) Streamlines for the case with LVGs

Fig. 41 Comparison of synergies between velocity and temperature fields for the case without and with LVGs.

In order to demonstrate the improvement of synergy between the flow field and temperature field, Fig. 41 shows comparison between the synergies between the flow and temperature fields for the cases without and with LVGs. Figure 41 (a) and (b) show the isotherms and streamlines for the case without LVGs. At the inlet of the heat exchanger, the isotherms and the streamlines are almost perpendicular to each other, which indicate that the synergy between the flow and temperature fields is very good. As flow continues to the wake zone, the isotherms are stretched and are parallel to the streamlines due to recirculation in the wake region. This means that the intersection angle between the velocity vector and the temperature gradient increases and the synergy between flow and temperature fields worsens. Figure 41 (c) and (d) show the isotherms and streamlines for the case with LVGs. Similar to the case without LVGs, the isotherms and the streamlines are almost perpendicular to each other at the inlet of the heat exchanger. As flow continues to the wake zone, the LVGs generated longitudinal vortices at the downstream of the oval tubes. The strong swirling secondary flow altered the local velocity and temperature fields that the intersection angle between the velocity and isotherms is increased. In other words, the angle between the velocity and the temperature gradient is decreased so that the synergy between velocity and temperature in the wake zone is improved and the overall heat transfer capacity of the heat exchanger is increased.

### (2) Effects of placements of the LVGs

In order to investigate the effects of locations of LVGs on the overall flow and heat transfer, the two structures shown in Fig. 42 are numerically investigated. The LVGs are either placed upper stream or downstream.

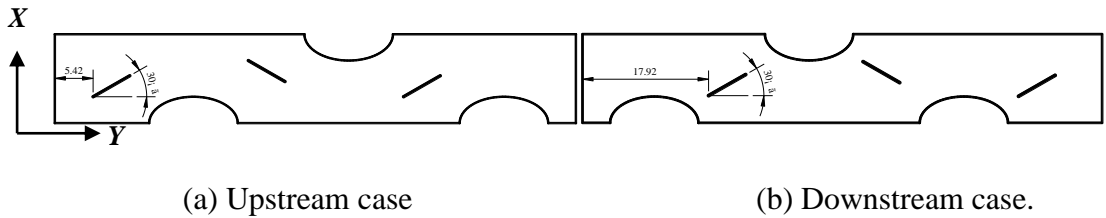

(a) Upstream case                           (b) Downstream case.
Fig. 42 Different strategies for placement of LVGs on fin-and-oval-tube heat exchangers

Figure 43 shows the average Nusselt number versus Reynolds number for the two different structures. It can be seen that the Nusselt numbers for both cases increase with increasing Reynolds number. In the ranges of Reynolds numbers studied, the heat transfer performance for the downstream case is better than that of the upstream case. At the entrance of the heat exchangers, the heat transfer performance is very good due to the entry effects. The synergy between the velocity and temperature fields is also very good so that the potential of heat transfer enhancement at the entrance region is very small. On the other hand, placement of LVGs at the downstream can effectively take advantage of the entry effects and is helpful to reduce the size of the wake zone to enhance heat transfer. This is also in agreement of the principle of synergy for heat transfer, i.e., enhancing heat transfer in the zone where synergy is poor.

Figure 44 shows the drag coefficient versus Reynolds number for the two different structures. It can be seen that the drag coefficients for both cases decrease with increasing Reynolds number. Under the same Reynolds number, the drag coefficient for the upstream case is slightly lower than that of the downstream case.

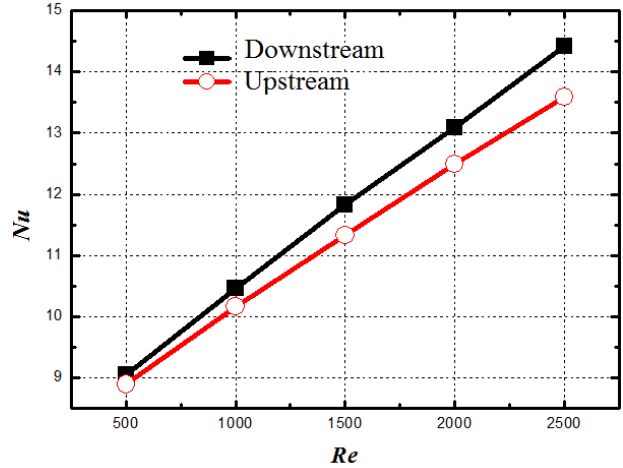

Fig. 43 Nusselt number vs. Reynolds number for different placements of LVGs

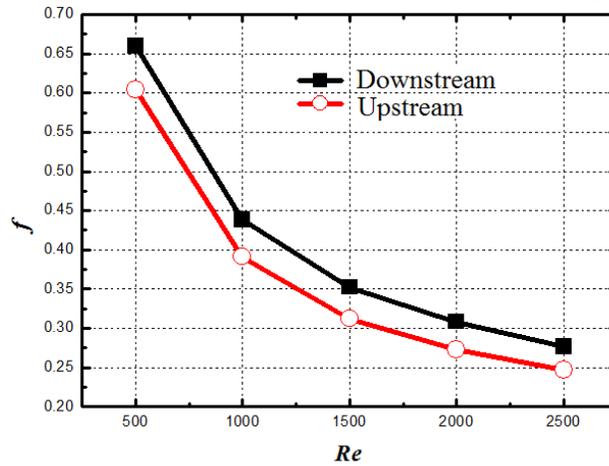

Fig. 44 Drag coefficient vs. Reynolds number for different placements of LVGs

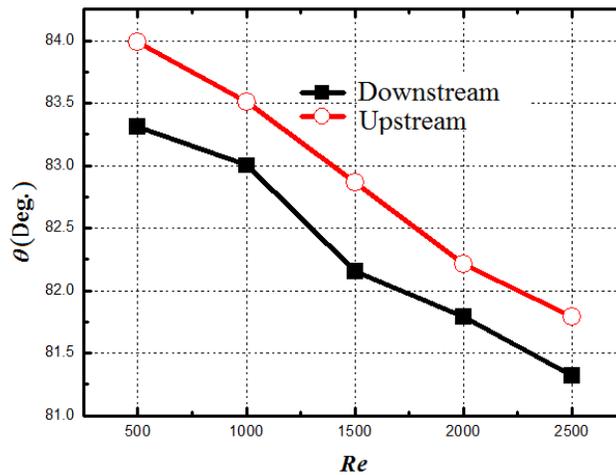

Fig. 45 Average intersection angle vs. Reynolds number for different placements of LVGs

Figure 45 shows the average intersection angle between velocity and temperature gradients versus Reynolds number for the two different structures. It can be seen that the placing the LVGs downstream can effectively improve the heat transfer in the wake zone where heat transfer is poorest and decrease the intersection angle. Consequently, heat transfer in the entire heat exchanger is enhanced.

In order to demonstrate variation of synergies between velocity and temperature fields for different LVG placements, Fig. 46 shows the isotherms and streamlines for the two different cases. The isotherms and streamlines for the upstream case are shown in Fig. 46 (a) and (b). As can be seen, the synergy at the entrance region is somehow improved due to LVGs. However, the synergy in the wake zone where heat transfer is poorest did not show much improvement. The isotherms and streamlines for downstream case are shown in Fig. 46 (c) and (d). The synergy at the entrance region is very good so that heat transfer enhancement in this region is not necessary. In the wake zone where heat transfer is poorest, the synergy is significantly improved, which, in turn, resulted in improvement of the heat transfer performance for the entire heat exchanger. Therefore, the downstream case is more helpful to improve heat transfer in the region where heat transfer is poorest and enhance heat transfer performance of the heat exchanger.

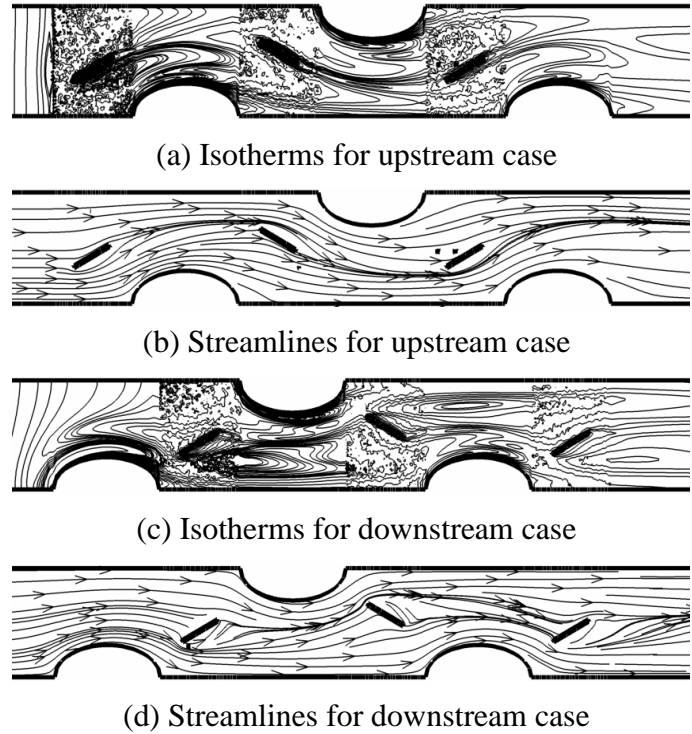

(a) Isotherms for upstream case

(b) Streamlines for upstream case

(c) Isotherms for downstream case

(d) Streamlines for downstream case

Fig. 46 Comparison between isotherms and streamlines for different placements of LVGs

**(3) Effects of angle of attack**

The effects of angle of attack for the downstream case are investigated. The number of rows are $n = 3$ and the range of Reynolds number is $Re = 500 \sim 2500$. The heat transfer enhancements at the following angles of attack are studied: $\alpha = 15°$, $30°$, $45°$, and $60°$ ($0°$ corresponds to the baseline case without LVGs).

Figure 47 shows the average Nusselt number of the heat exchanger with LVGs versus Reynolds number at different angles of attack. The effect of longitudinal vortices on heat transfer enhancement not only dictated by the intensity of the vortices, but also depends on the persistency of the vortices. Both intensity and persistency of the longitudinal vortices are affected by the angle of attack α. It can be seen from the figure that first the average Nusselt number increases with the increasing angle of attack α, then the average Nusselt number reaches the maximum at the angle of attack of α = 30°, and finally the average Nusselt number decreases with increasing angle of attack. When α < 30°, the intensities of the longitudinal vortices increase with increasing angle of attack; therefore, the average Nusselt number increases. In fact, the LVGs not only generate longitudinal vortices, but also generate some transverse vortices, which can also enhance heat transfer just like the longitudinal vortices. The transverse vortices can affect the persistency of the longitudinal vortices, or even destroy the longitudinal vortices. When α > 30°, although the intensity of the longitudinal vortices continuous to increase, the persistency of the longitudinal vortices is decreased due to effects of transverse vortices. Therefore, the results of heat transfer enhancement are affected and the average Nusselt number is slightly decreased. When the angles of attack continuous to increase to beyond 65°, what generated by LVGs are mainly transverse vortices and the longitudinal vortices are destroyed. Consequently, the effectiveness of LVGs on heat transfer enhancement is significantly affected.

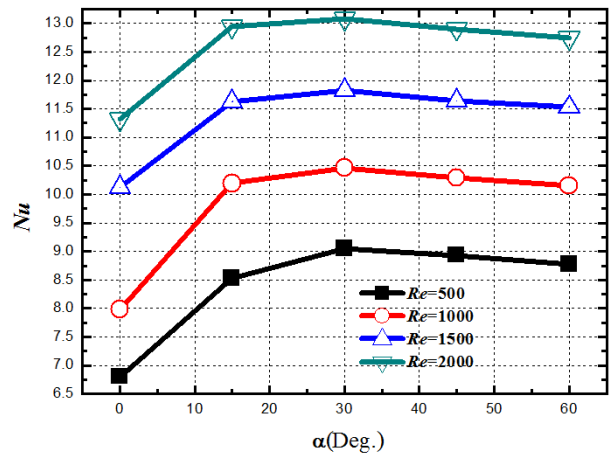

Fig. 47 Nusselt number vs. Reynolds number at different angle of attack

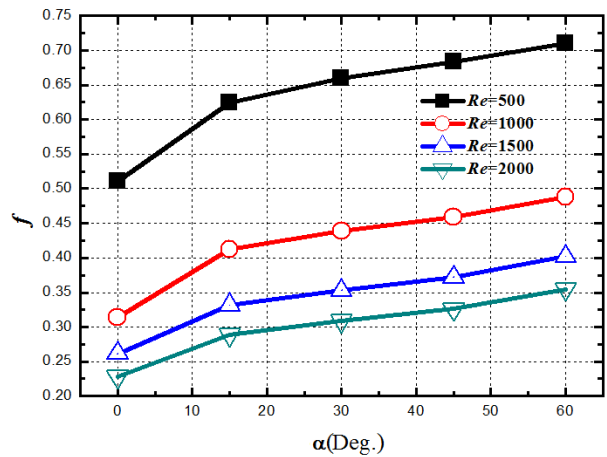

Fig. 48 Drag Coefficient vs. Reynolds number at different angle of attack

Figure 48 shows the drag coefficient versus Reynolds number at different angles of attack. It can be seen that the drag coefficient increases with increasing angle of attack. The reason is that the as angle of attack increases, the form drag increases and it causes increased pressure loss and higher drag coefficient.

Figure 49 shows the comparison of synergies between velocity and temperature fields for the cases with angles of attack of 30° and 60°. At the entrances of the two heat exchangers and before the LVGs, the isotherms and streamlines for the two cases are almost the same: the streamlines and isotherms are almost perpendicular to each other, i.e., the synergies for both heat exchangers are very good. As the fluids pass the LVGs, the synergy between the temperature and velocity fields in the wake zone for the case with an angle of attack of 30° is significantly improved. On the contrary, the angle between the isotherms and streamlines in the wake zone for the case with an angle of attack of 60° is very small, and the isotherms and streamlines are even parallel in some region. Therefore, the synergy between the temperature and velocity fields is poor, and there is no significant improvement on synergy for the case of with an angle of attack of 60°.

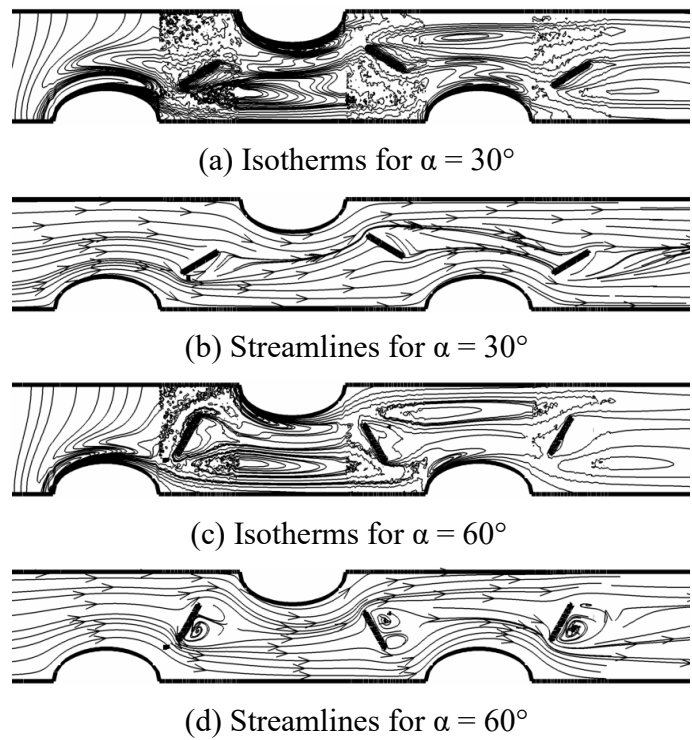

(a) Isotherms for α = 30°

(b) Streamlines for α = 30°

(c) Isotherms for α = 60°

(d) Streamlines for α = 60°

Fig. 49 Comparison of synergies between temperature and velocity fields at different angles of attack

### (4) Effects of the number of rows of tubes

Based on the conclusion that the angle of attack of α = 30° can provide the best result on heat transfer enhancement, the effect of the number of rows on the performance of the fin-and-oval-tube heat exchanger with delta winglets is studied. The numbers of rows are 2, 3, 4 and 5, and the angle of attack of the downstream LVGs are 30°. The range of the Reynolds number is Re = 500 ~ 2500.

Figure 50 shows the average Nusselt number of the fin-and-oval-tube heat exchanger with LVGs versus the number of rows of tubes at different Reynolds numbers. It can be seen that the Nusselt numbers decrease with increasing number of rows of tubes, and trends for different Reynolds number are the same. The variations of the average Nusselt number from $n = 2$ to $n = 3$ is more significant than that from $n = 3$ to $n = 4$ or from $n = 4$ to $n = 5$.

Figure 51 shows the drag coefficients of the fin-and-oval-tube heat exchanger with LVGs versus the number of rows of tubes at different Reynolds numbers. It can be seen that the drag coefficient decreases with increasing number of rows of tubes, and trends for different Reynolds number are similar to each other. The variations of the drag coefficients from $n = 2$ to $n = 3$ is more significant than that from $n = 3$ to $n = 4$, which is significant than that from $n = 4$ to $n = 5$. Thus, one can conclude that, as the number of rows increases, the effects of the number of rows on the average Nusselt number and drag coefficient become less significant.

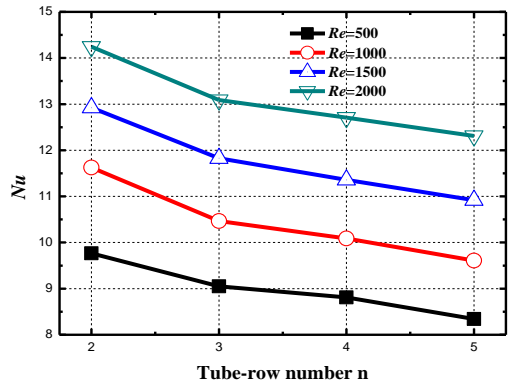

Fig. 50 Average Nusselt number vs. number of row of tubes

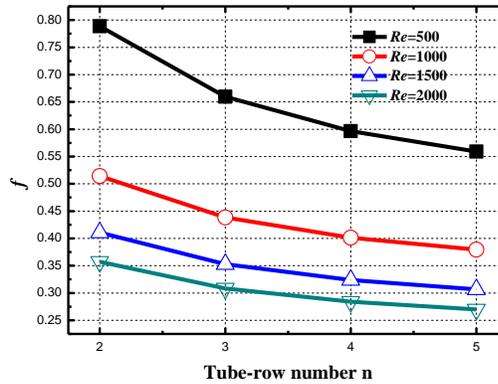

Fig. 51 Drag coefficient vs. number of row of tubes

Figure 52 shows the average intersection angle between the temperature and velocity fields versus number of row of tubes at different Reynolds numbers. As the number of rows of tubes increases, the average intersection angle increases. The variations of the intersection angle from $n = 2$ to $n = 3$ is more significant than that from $n = 3$ to $n = 4$, which is significant than that from $n = 4$ to $n = 5$. In other words, the intersection angle increases with increasing number of row of tubes but the increase becomes less significant at large number of rows; this is consistent with the variation of Nusselt number shown in Fig. 50. The variations of intersection angle at different Reynolds numbers are similar to each other.

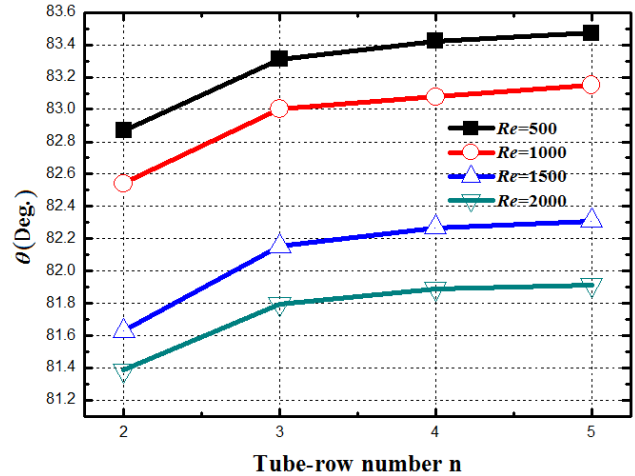

Fig. 52 Average Nusselt number vs. number of row of tubes

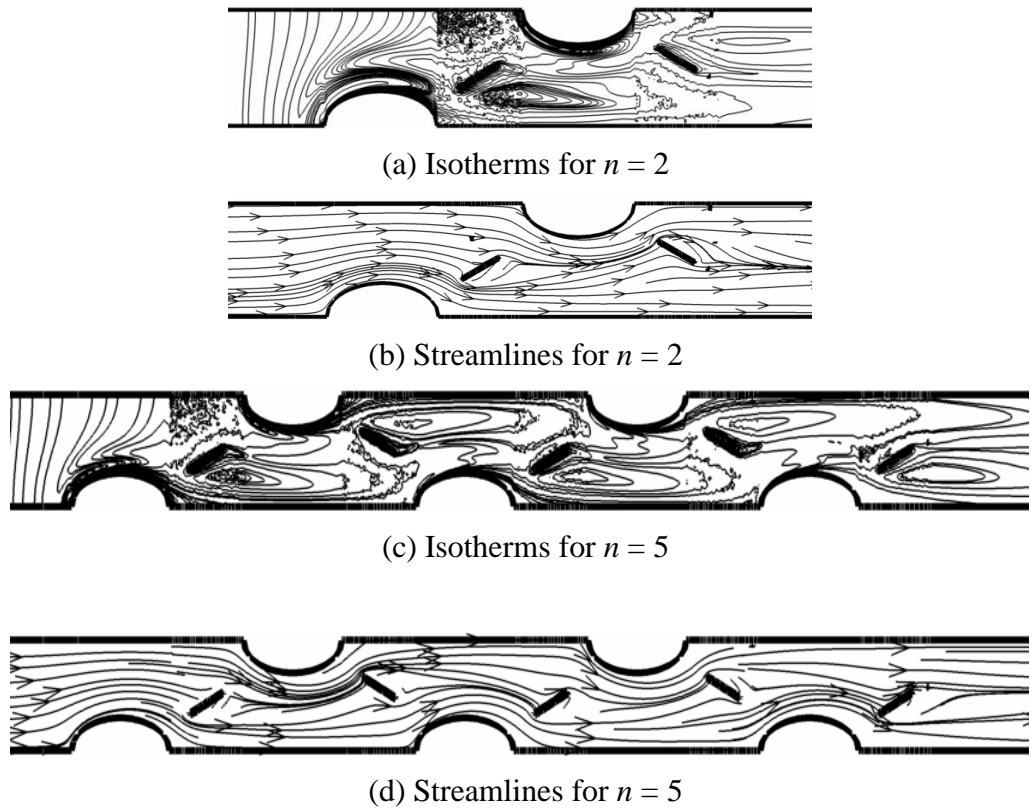

(a) Isotherms for $n = 2$

(b) Streamlines for $n = 2$

(c) Isotherms for $n = 5$

(d) Streamlines for $n = 5$

Fig. 53 Comparison for Synergies between temperature and velocity fields

Figure 53 shows the comparison of synergies of temperature and velocity fields at different number of rows of tubes. At the entrances of the two heat exchangers, the isotherms and streamlines for both cases are almost perpendicular to each other, i.e., the synergies for both heat exchangers are very good. As for the isotherms and streamlines for the case of $n = 5$, the isotherms and streamlines before the second row are almost identical to case of $n = 2$. As the flow further develops, the isotherms for the case of $n = 5$ gradually stretch and their angle with streamlines become smaller and smaller. In other words, the synergy between the temperature and velocity field becomes poorer and poorer. As a result, the synergy for the case of $n = 5$ is worse than that of the case of $n = 2$.

### 3.2.4 Application of LVGs to fin-and-tube heat exchangers with multiple row of tubes

As mentioned earlier, the two most commonly reported LVG placement strategies are "common-flow-down" and the "common-flow-up" approaches [10], which can be abbreviated as CFD and CFU approaches (see Fig. 54). Most existing researches about heat transfer enhancement via LVGs deal with the CFD approach, while the research on flow and heat transfer in CFU approach is scant.

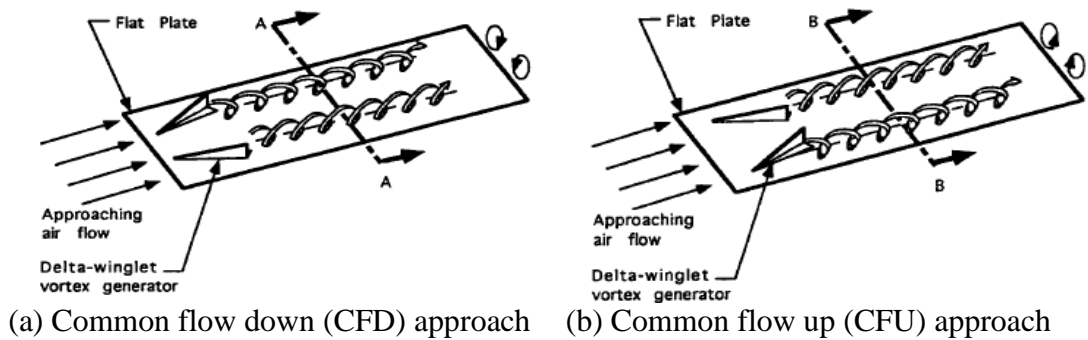

(a) Common flow down (CFD) approach    (b) Common flow up (CFU) approach
Fig. 54 Schematic of different placement strategies of LVGs

Under CFD approach, the fluid between a pair of delta (or rectangular) winglet flows toward the fin with LVGs so that it is called flow *down*. CFD is characterized by the fact the distance between the heads of the delta winglets are longer than that between the tails. On the other hand, for the CFU approach, the fluid between a pair of delta (or rectangular) winglet flows away from the fin with LVGs so that it is called flow *up*. The distance between the heads of the delta winglets are shorter than that between the tails.

With regard to the studies on the heat transfer enhancement by LVGs with CFU placement, Joardar and Jacobi[10] numerically simulated fin-and-tube heat exchangers with 7 rows of tubes. They found that at Re = 850, heat transfer for the case of heat exchangers with 3 rows of tubes arranged inline and delta winglets is enhanced by 32.3% and the corresponding drag coefficient is increased by 41%. They also obtained another interesting result that when the tube arrangement is changed to staggered, the same heat transfer enhancement can be obtained while the drag coefficient was only increased by about 33%. However, no explanation about the effect of tube arrangement on the drag coefficient was provided. Joardar and Jacobi[44] performed experimental investigation for fin-and-tube heat exchanger with delta winglet. They found that when the Reynolds number was between 220 to 960, one row of delta winglets could increase heat transfer coefficient by 16.5 ~ 44% and the corresponding increase of drag coefficient was less than 12%. For the case of three rows of delta winglets, the heat transfer is enhanced by 29.9 ~ 68.8%，and the corresponding drag coefficient is increased by 26 ~ 87.5%。Torri et al. [8] experimentally studied flow and heat transfer of fin-and-tube heat exchanger with only one row of delta winglets. For staggered tube arrangement, the Colburn factor was increased by 10 ~ 30%, while the drag coefficient, *f*, is decreased by 34 ~55%. For the case of inline arrangement, the Colburn factor was increased by 10 ~ 20%, and the drag coefficient, *f*, is decreased by 8 ~ 15 %. Kwak and Torri[9] studied flow and heat transfer of fin-and-tube heat exchangers with inline and staggered arrangements and one or two rows of delta winglets. They obtained the similar conclusion that the CFU placement of LVGs enhanced heat transfer while decreased drag.

Most existing researches about LVGs with CFU placement used delta winglets while the researches on rectangular winglets are rare. The reason that the CFU placement has advantages is that in addition to generate strong secondary swirling flow, a convergent flow channel can be formed between the LVG and the tube wall. The fluid is accelerated when it passes the convergent channel, and impinging to the tubes in the next row to enhance heat transfer. This high-velocity impinging flow can also delay the separation of the boundary layer and reduce the size of the wake zone so that the form drag of the tube banks can also be reduced. When rectangular winglets are employed, these advantages are more significant since more fluids are accelerated as they pass the convergent channel. We investigated the applications of rectangular winglets on enhancement of fin-and-tube heat exchangers and performed parametric studies [45].

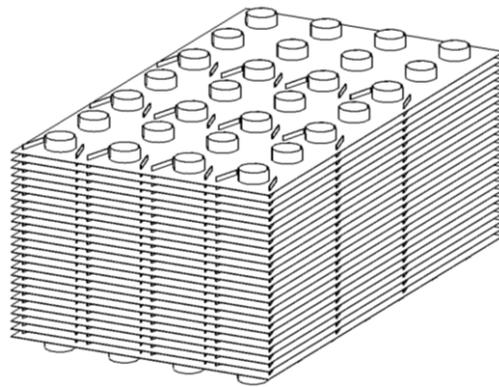

Fig. 55 Schematic of the core region of a fin-and-tube heat exchanger with rectangular winglets

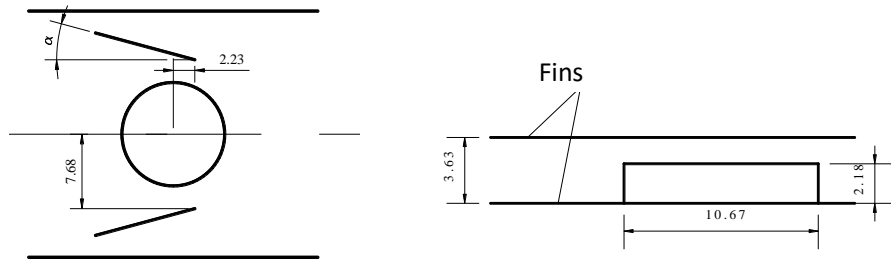

Fig. 56 Dimensions and the placement of LVGs with respect to the tube (unit: mm)

The rectangular winglets are placed on the two sides of the tube using CFU approach and convergent flow channels are formed between LVGs and the tube wall. The reason that rectangular winglets, instead of delta winglets, are adopted was that more fluid can pass the convergent channel that is similar to a convergent nozzle. The accelerated fluid impinges to the tubes in the next row. The impingement of the fluid reduces the boundary layer thickness and increases the temperature gradient, which ultimately enhance heat transfer. Therefore, the LVGs with CFU placement enhance heat transfer by the combined effects of longitudinal vortices and fluid impingement. The fluid accelerated by the convergent channel not only can enhance heat transfer by impingement, but can also delay the boundary lay separation. Together with the strong swirling secondary flow, the fluid accelerated by the convergent channel can decrease the size of the wake zone to reduce the form drag of the tube. Figure 55 shows the schematic of the core region of a fin-and-tube heat exchanger with rectangular winglets, while Fig. 56 shows the winglet type vortex generator dimensions and the placement with respect to the tube.

**(1) Effect of the angle of attack of the LVGs**

Figure 57 shows the different configurations for fin-and-tube heat exchangers without and with rectangular winglets. Four different angles of attack are studied: $\alpha = 0°$, $10°$, $20°$, and $30°$ ($\alpha = 0°$ corresponds to the baseline case without LVGs).

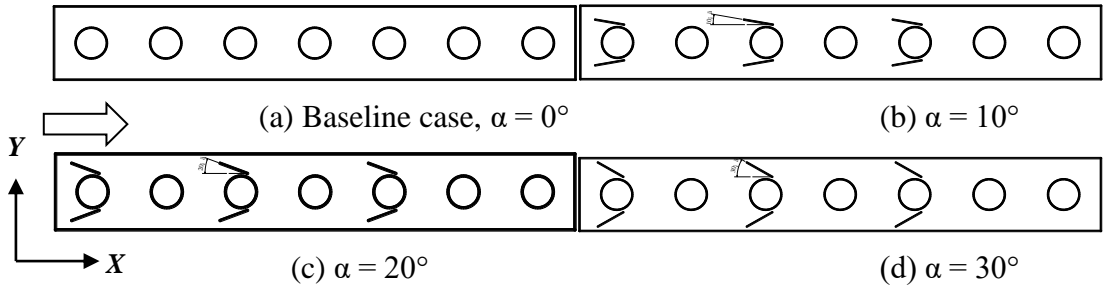

Fig. 57 Different configurations for fin-and-tube heat exchangers with and without rectangular winglets

Figure 58 shows the velocity and streamlines distributions at the middle plane perpendicular to the tubes. Due to the periodical feature, the velocity and streamline distributions for the tube rows downstream are similar to that of the first two rows. In order to show the detail of the fluid fields and save space, only the distributions for the first two rows are shown. It can be seen from Fig. 58(a) that the velocity at the tail of the tube is very low. The existences of the streamline in the wake region indicate that recirculation takes place in the wake region. The axis of the recirculation flow is perpendicular to the main flow direction so that this circulation flow is called transverse vortices, which are generated by separation of the fluid from the tube. The transverse vortices independently swirl in the wake zone and almost do not exchange mass with the air flow and is almost isolated from the main flow. On the other hand, it can be seen from Fig. 58(b)-(d) that the streamlines are stretched in the middle of the wake zone. This constrict streamlines are formed due to generation of longitudinal vortices that induce high-momentum fluid to the wake zone and decreases the size of the wake zone. As the angle of attack increases, the longitudinal vortices intensify and the streamlines are even more curved in the central region of the wake zone; the size of the wake zone decreases even more. We also found that the LVGs had very little on the heat transfer in the wake zone of the tubes in the next row.

Figure 59 shows the temperature contour at the middle-plane perpendicular to the tube at Re = 850. For the case without LVGs (Fig. 59(a)), the temperature gradient in the wake zone of the tube is very small. For the cases with LVGs as shown in Fig. 59(b) ~ (d), the temperature gradient gradually increases with increasing angle of attack. Since the wake zone is usually the poorest region for heat transfer in the entire heat exchanger, increasing temperature gradient in this region results in increase of heat flux and ultimately improves heat transfer in this region. After the fluid flows through the convergent narrow and long channel, it impinges the next tubes in the next row and decreases their boundary thicknesses. The temperature gradient increases, and the heat flux increases, which is very helpful for heat transfer enhancement. As the angle of attack increases from 0° (baseline case) to 30°, the temperature gradient in the region before the tubes in the second row gradually increases because the fluid impingement velocity increases with increasing angle of attack due to increased ratio of the cross-sectional area of the convergent channel. This resulted in thinner boundary layer in front of the tubes in the second row and consequently higher temperature gradient.

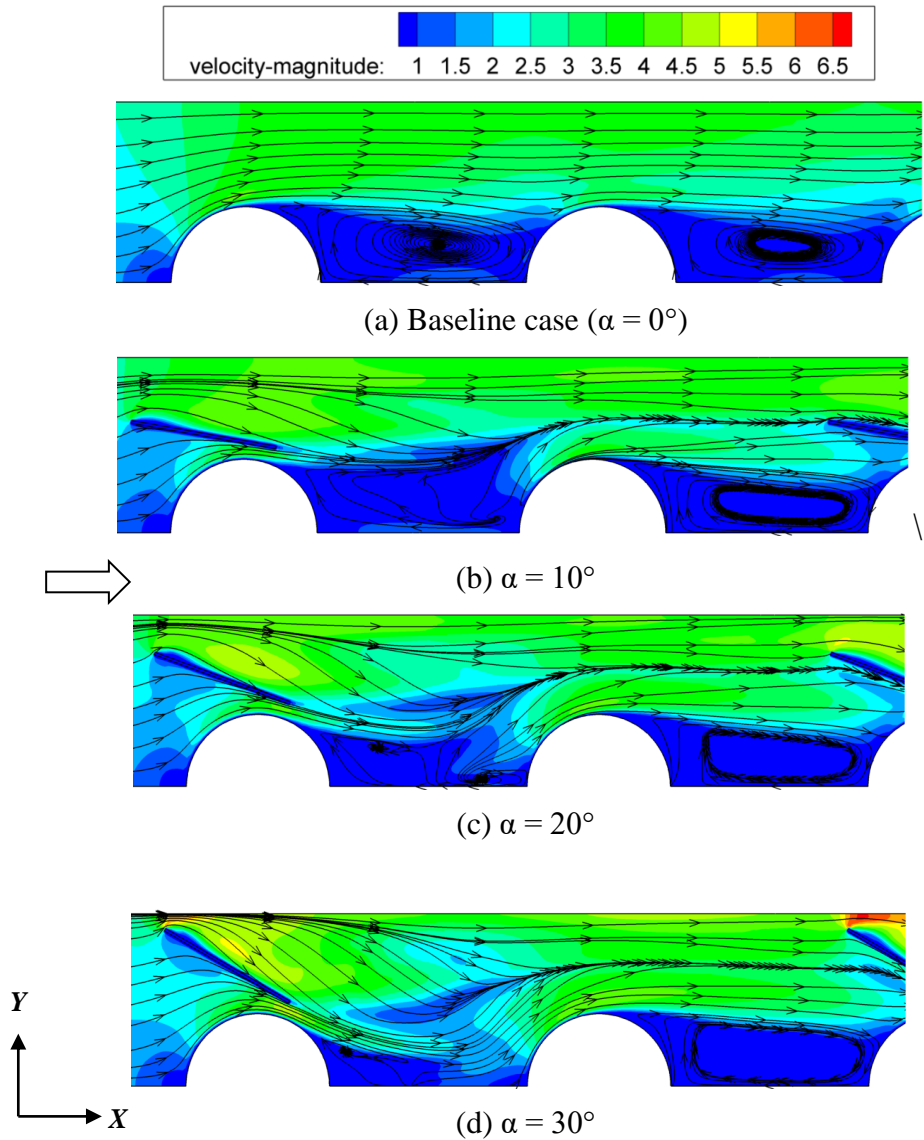

(a) Baseline case (α = 0°)

(b) α = 10°

(c) α = 20°

(d) α = 30°

Fig. 58 Velocities and streamlines distribution at different angles of attack (Re = 850)

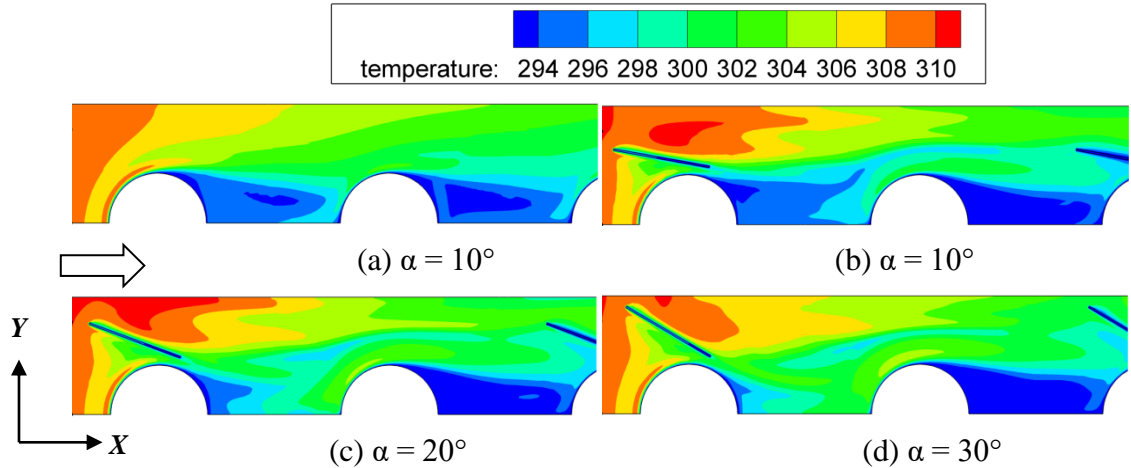

Fig. 59 Temperature distribution at different angles of attack (Re = 850)

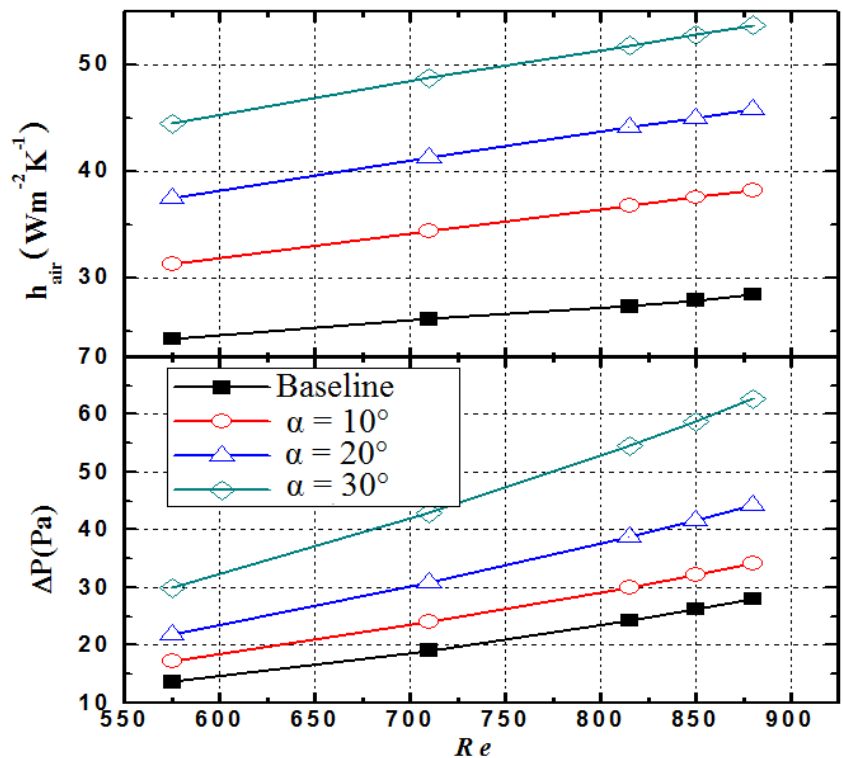

Fig. 60 Heat transfer coefficient and pressure drop vs. Reynolds number

Figure 60 shows the average heat transfer coefficient and drag coefficient versus Reynolds number at different angles of attack. As can be seen from Fig. 60(a), the heat transfer coefficient increases with increasing Reynolds number since the boundary layer thickness decreases with increasing Reynolds number and the mixing between the hot and cold fluids is promoted by LVGs. Compared to the baseline case and within the range of the Reynolds numbers that was studied (Re = 575 ~ 880), the heat transfer coefficient on the outside of the tubes is increased by 28.8-34.5%, 54.5-61.5% and 83.3-89.7% for the angles of attack of 10°, 20°, and 30°, respectively. These results indicate that the rectangular winglets can effectively increase the heat transfer coefficient on the outside of the tube. As can be seen from Fig. 60(b), enhancement of heat transfer is always accompanied by increase of drag. Compare to the baseline case and within the range of the Reynolds numbers that was studied (Re = 575 ~ 880), the drag coefficient is increased by 21.9~26.9%, 58.1~61.9% and 119.2~125.3% for the angles of attack of 10°, 20°, and 30°, respectively. The reason that the drag coefficient increases with increasing angle of attack is that the form drag of the LVGs increases with increasing angle of attack.

Figure 61 shows the overall performance of the heat exchanger versus Reynolds number at different angles of attack. In the range of Reynolds number studied (Re = 575 ~ 880), $j/f$ for the case of $\alpha = 10°$ is increased by 1.4~10.3% compared to the baseline case. For the case of $\alpha = 20°$, the performance of the heat exchanger with LVGs is worse than the baseline case when the Reynolds number is below 800; $j/f$ for the case with LVGs is higher than the baseline case if the Reynolds number is above 800. Effects of Reynolds number on the overall performance can be explained by its different effects on pressure drops across tube and LVGs. As Reynolds number increases, the fluid accelerated by the convergent narrow and long channel achieves higher velocity so that the size of the wake region is reduced and the pressure drop across tubes is decreased. On the other hand, the pressure drop due to LVGs increases with increasing Reynolds number. At higher Reynolds number, decrease of the pressure drop across tubes becomes dominant so that $j/f$ for the case with LVGs at higher Reynolds number is higher than that of the baseline case. For the case of $\alpha = 30°$, the increase of pressure drop due to LVGs for the entire range of Reynolds number is higher than the decrease of pressure drop across the tubes so that the $j/f$ for the case with LVGs is always lower than that of the baseline case. In the range of Reynolds number that was studied, the overall performance measured by $j/f$ is 15.5~17.0% lower than that of the baseline case.

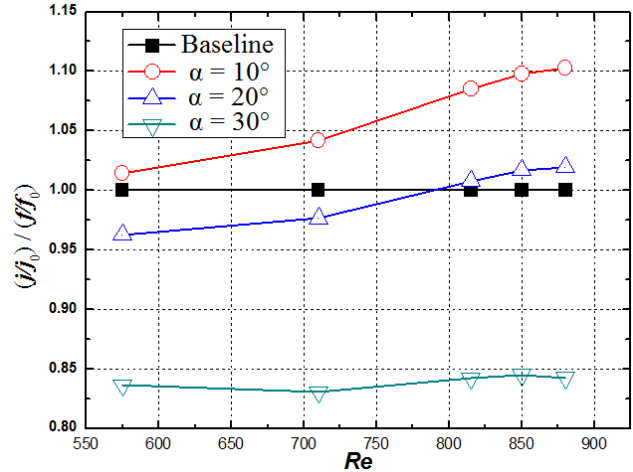

Fig. 61 Overall performances of heat exchangers vs. Reynolds number

It can be concluded from the above discussion that among the four different structures, the heat exchanger with LVGs at angles of attack of 10° and 20° are better than the other two cases. The heat exchanger with LVG at $\alpha = 10°$ could enhance heat transfer by 30% in comparison to the baseline case while the increase of pressure drop is very small. However, this enhancement does not show any significant advantage over other traditional heat transfer enhancement techniques. On the other hand, the heat exchanger with LVGs at angle of attack of 20° enhances heat transfer by as much as 60% and the heat transfer enhancement at higher Reynolds number (Re > 800) is higher than the increase of pressure drop, i.e., $(j/j_0)/(f/f_0) > 1$, which is the most challenging task in heat transfer enhancement. Therefore, the following parametric studies will be carried out for the fin-and-tube heat exchangers with an angle of attack of 20°.

**(2) Effects of the number of LVGs**

Figure 62 shows the structures of fin-and-tube heat exchanger with different number of LVGs, which are rectangular winglet pairs (RWPs). The angle of attack is fixed at 20°, the range of Reynolds number is Re = 575~880, and the number of the row of tubes is 7.

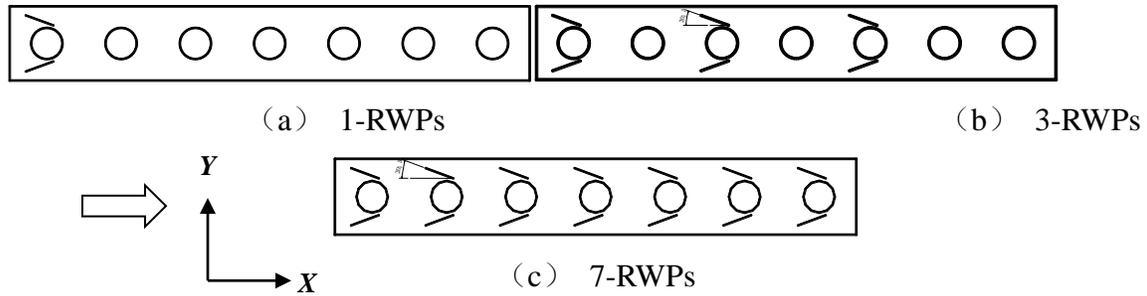

（a） 1-RWPs  （b） 3-RWPs

（c） 7-RWPs

Fig. 62 Fin-and-tube heat exchangers with different numbers of LVGs (α = 20°)

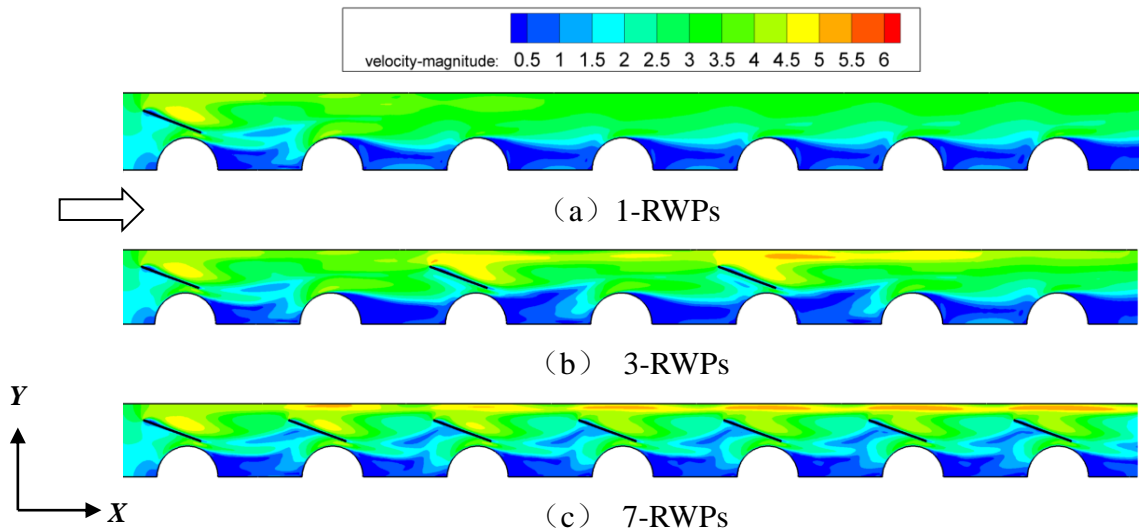

（a） 1-RWPs

（b） 3-RWPs

（c） 7-RWPs

Fig. 63 Distributions of the velocity magnitudes in fin-and tube heat exchangers

Figure 63 shows the distribution of the velocity magnitudes in fin-and tube heat exchangers with different number of LVGs. The high-momentum (or high-velocity) regions can be observed from all 3 different structures. As it was discussed before, when the fluid passes the LVGs, part of it becomes high-velocity swirling secondary flow and part of it is accelerated by the convergent narrow and long flow channel. For the case with one RWP the fluid is accelerated only near the first row of tubes, while for the case with three RWPs the fluid is accelerated near the 1$^{st}$, 2$^{nd}$, and 5$^{th}$ rows of the tubes. For the case of seven RWPs, the fluid is accelerated near every row of tubes and the velocity at the downstream of the LVGs gradually increases as the flow gradually develops. It can also be seen from Fig. 63 that due to the strong swirling secondary flow and acceleration by the convergent channel, the size of wake zone behind the first row of the tube for the case of one RWP is significantly decreased. For the case of three RWPs, the size of wake zones behind the $1^{st}$, $3^{rd}$, and $5^{th}$ tubes are significantly decreased. When the number of RWPs increases to seven, the sizes of the wake zones behind all seven rows of tubes are decreased. Due to the periodic structure of the flow channel, the wake zones behind all seven rows of tubes are almost identical.

Figure 64 shows the temperature contours in the fin-and-tube heat exchangers with different numbers of LVGs. It can be seen that the temperature contours for three structures are almost the same before the LVGs. The temperature gradient behind the first row of the tube for the case of one RWP is higher than that behind the rest of the tubes. For the case of three RWPs, the temperature gradients behind the $3^{rd}$ and $5^{th}$ rows of the tubes is higher than that corresponding to the case of one RWP. Comparison between the cases of three and seven RWPs indicates that the temperature gradient behind every tube for the case of seven RWPs is higher than that for the case of three RWPs. The reason for this difference is that the wake zone behind every (except the $7^{th}$) tube in the case of seven RWPs is affected by both LVGs located at upstream and downstream. On the contrary, the wake zone behind every tube in the case of three RWPs is affected by only one RWP. Therefore, the wake zones for the case of 7 RWPs are disturbed more and the temperature gradients in the wake zones are also higher.

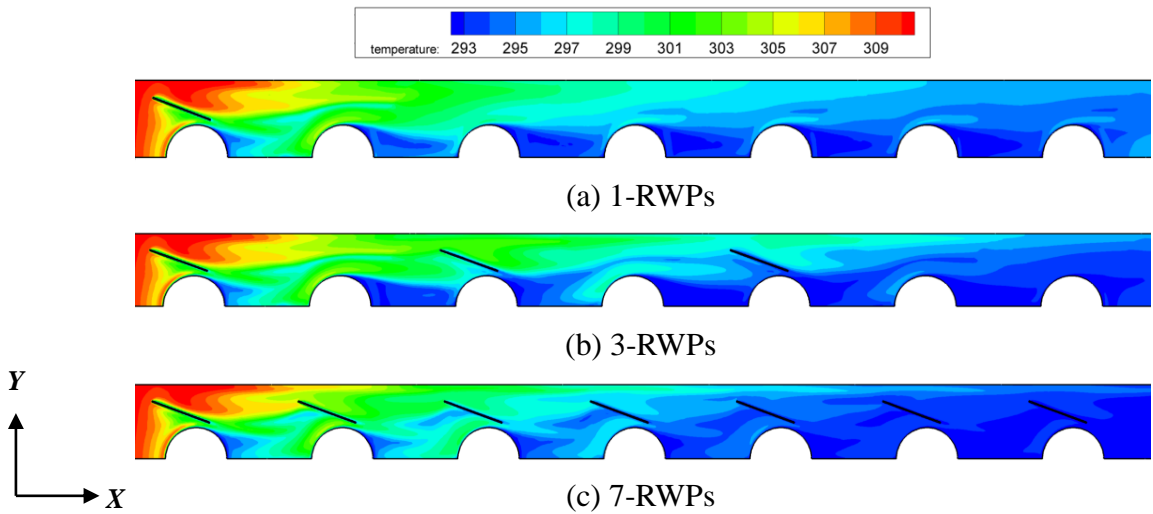

(a) 1-RWPs

(b) 3-RWPs

(c) 7-RWPs

Figure 64 temperature contours in the fin-and-tube heat exchangers

Figure 65 shows the effect of the number of LVGs on the flow and heat transfer. Comparison between the air side heat transfer coefficients for different number of LVGs is given in Fig. 65(a). Compared to the baseline case, the heat transfer coefficients were respectively increased by 22.7~25.5%, 54.6~61.5 and 87.5~105.1% for the cases of one, three and seven RWPs. While more LVGs result in more significant heat transfer enhancement, the pressure drop also increases with increasing number of LVGs, as indicated by Fig. 65(b). Compared to the baseline case, the pressure drop were respectively increased by 22~24.5%, 58.1~62 and 123~127.6 for the cases of one, three and seven RWPs.

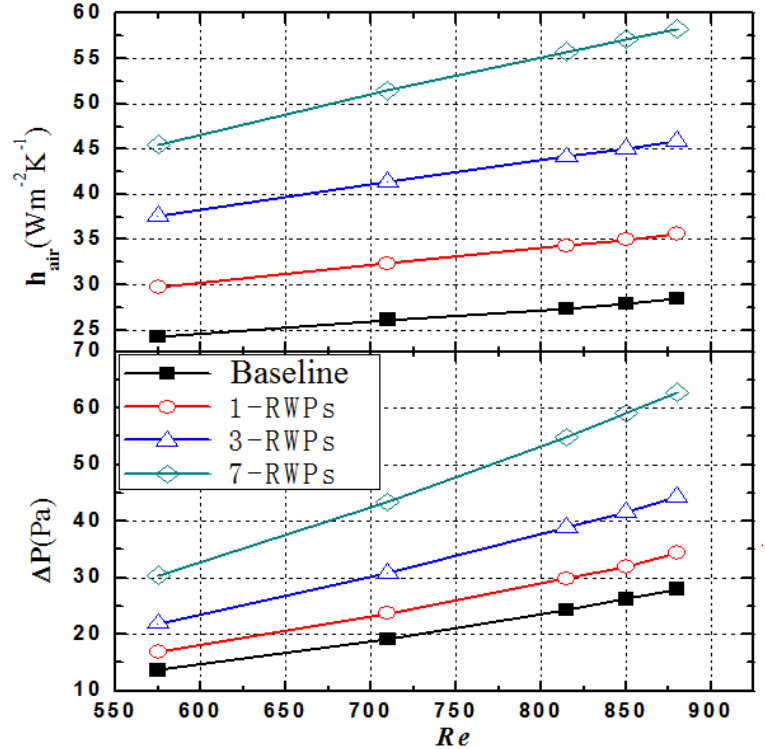

Fig. 65 Heat transfer coefficient and pressure drop vs. Reynolds number

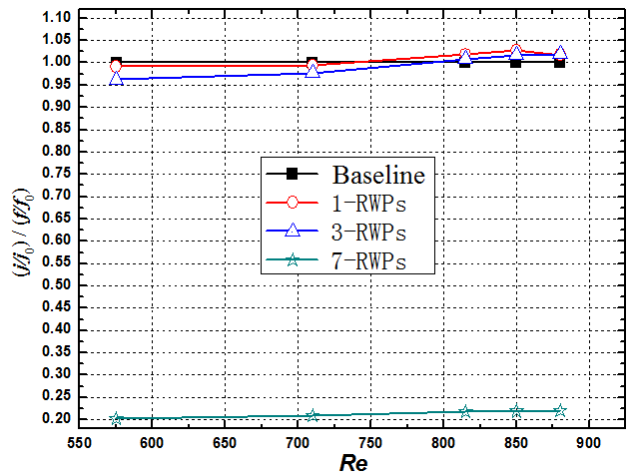

Figure 66 Overall heat transfer performance versus Reynolds number at different LVG numbers

Figure 66 shows the overall heat transfer performance versus Reynolds number at different number of LVGs. It can be seen that for the case of seven RWPs the overall heat transfer performance $j/f$ in the entire range of Reynolds number is lower than that of the baseline structure. For the cases of one and three RWPs, the overall heat transfer performance $j/f$ is lower than that of the baseline structure when Reynolds number is under 815, but higher than that of the baseline structure when Reynolds number exceeds 815. Compared to the baseline case, the overall heat transfer performances for the cases of one and three RWPs are increased by 1.7~2.7% and 0.7~2.0, respectively.

**(3) Effects of placement of LVGs**

Figure 67 shows the schematic of inline and staggered arrangements of LVGs. For the case of inline arrangement of LVGs (see Fig. 67(b)), three pairs (6 total) of rectangular winglets are symmetrically placed on the two sides of the 1$^{st}$, 3$^{rd}$, and 5$^{th}$ tubes. For the case of staggered arrangement of LVGs, the 6 rectangular winglets are alternatively placed on the up and down sides of the tubes 1 to 6. The numbers of rectangular winglets for both arrangements are the same and the angle of attack is $\alpha = 20°$. The range of Reynolds number is 575 ~ 880 and the number of rows is 7.

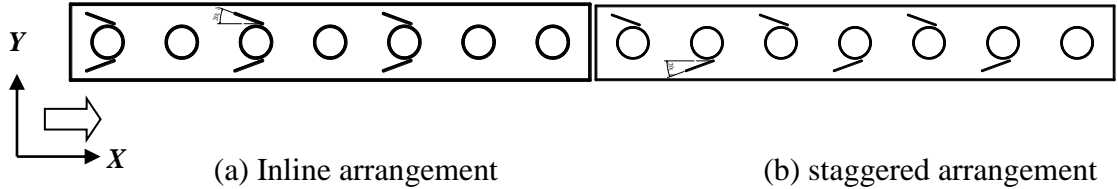

(a) Inline arrangement  (b) staggered arrangement
Fig. 67 Inline and staggered arrangements of LVGs

Figure 68 shows the comparison of the contours of the magnitudes of velocities for inline and staggered arrangements of LVGs. For the case of inline arrangement, the velocity is symmetric since the geometric structure of the flow channel is symmetric. Due to the strong swirling flow induced by the LVGs and the impingement of the accelerated flow, the sizes of the wake zones behind the 1$^{st}$, 3$^{rd}$, and 5$^{th}$ tubes are somewhat decreased. Since the LVGs are symmetrically placed on the two sides of the tubes, two symmetric high-velocity jets on the two sides of the tube can be formed. However, the velocity components of the two jets in the transverse directions are in the opposite direction so that these two symmetric jets can cancel each other's transverse velocity components. Consequently, the effects of these two jets on the wake zones are weakened and the heat transfer enhancements due to the high-velocity jets are also weakened. On the other hand, the velocity distribution for the case of the staggered LVGs is not symmetric because its geometric structure is not symmetric. Since the six rectangular winglets are staggered on the sides of the 1$^{st}$ to 6$^{th}$ tubes, every LVG can independently improve the heat transfer in the wake zone behind the tube to its full potential without affected by any other LVGs. While the heat transfer in the wake zones of only three tubes are improved for the case of inline arrangement, the heat transfer for the first 6 tubes are improved at certain degree for the case of staggered arrangement.

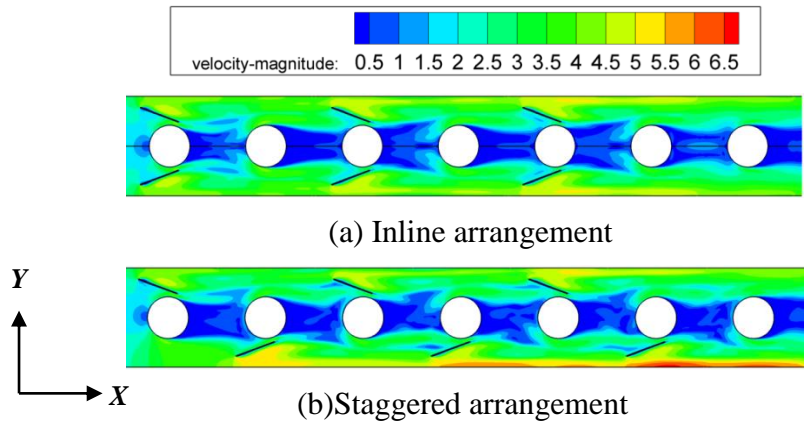

(a) Inline arrangement

(b)Staggered arrangement

Fig. 68 Comparison of velocity magnitudes for different arrangements of LVGs (unit: m/s; Re = 850)

Figure 69 shows the comparison of the temperature contours for different arrangements of LVGs. For the case of inline arrangement, the temperature is symmetric. The velocities in the wake zones are very low and their mass and energy exchanges with the main flow region are also very low. Therefore, the fluid temperatures in the wake zones are very close to the tube wall temperatures so that the temperature gradients and heat flux are very low in the wake zones. Due to strong swirling secondary flow and jet impingement effects, the sizes of the wake zones behind the $1^{st}$, $3^{rd}$, and $5^{th}$ tubes are significantly decreased. The temperature gradients are increased in these wake zones, which, in turn, helps heat exchange between the tube and the fluid. For the $2^{nd}$, $4^{th}$ and $6^{th}$ tubes where no LVGs on their sides the low-heat flux zone extended all the way to the next tube, which degrade the heat transfer performance. For the case of staggered arrangement of LVGs, the temperature distribution is not symmetric. Due to the effects of swirling secondary flow and jet impingement, the sizes of the wake zones behind the $1^{st}$ through $6^{th}$ tube are decreased somewhat and the low-heat flux zones becomes smaller; all of these facts attributed to heat transfer enhancement.

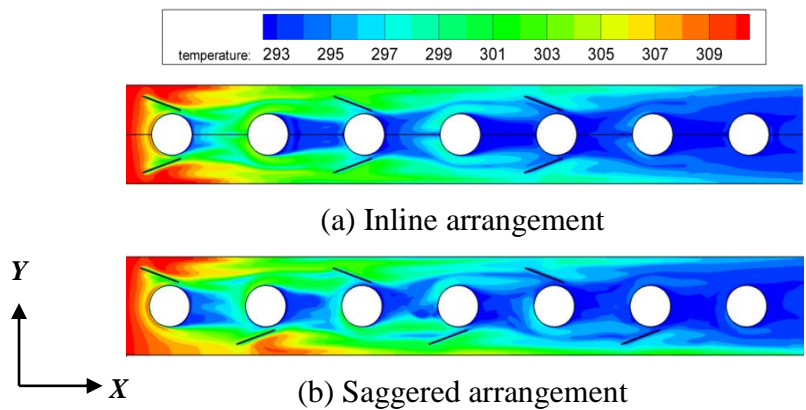

Fig. 69 Comparison of the temperature contours for different arrangements of LVGs.

Figure 70 shows the comparison of flow and heat transfer for different arrangements of LVGs. The air-side heat transfer coefficients versus Reynolds number is shown in Fig. 70(a). Compared to the inline arrangement, the air side heat transfer coefficient for staggered arrangement of LVGs is 0.5~2.5% higher. Meanwhile, the pressure drop for staggered arrangement of LVG is 4.5~8.3% lower. Therefore, the staggered arrangement allows further reduction of pressure drop while keeps the level of heat transfer enhancement unchanged, which can be attributed to the asymmetric placement of LVGs. It can be expected that as Reynolds number further increases, the asymmetric structure will further reduce the pressure drop.

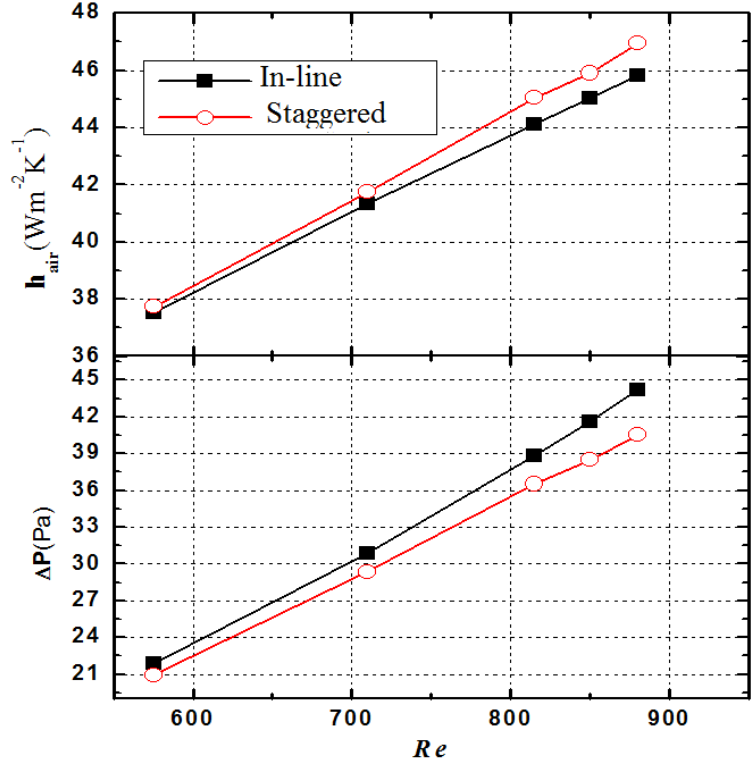

Fig. 70 Comparison of heat transfer and pressure drop for different placement of LVGs

## 4 Conclusions and Outlooks

Compared to the traditional heat transfer enhancement techniques, the LVGs have unparalleled advantages. While the capability of heat transfer is significantly enhanced, the pressure drop only increases by a small degree, or even decreases in some cases. It can achieve the toughest criterion of $(j/j_0)/(f/f_0)>1$ that is very difficult to achieve by the traditional measures.

The heat transfer enhancement by LVGs is affected by many factors, such as angle of attack, aspect ratio, locations, arrangement, and flow patterns. Different optimized structures exist depending on the different operating conditions. Since the parameter space for the optimization of LVGs is huge, experiment alone is difficult to obtain the optimized design. In addition, the in-depth mechanisms of the heat transfer can only be revealed by analyzing flow field, temperature and pressure distributions. Therefore, it is imperative to use numerical approaches to carry out studies on design optimization of heat exchangers.

Based on the above very detailed review on the state-of-the-art researches on heat transfer enhancements by LVGs, we believe that the following three topics will possibly become the "hot spots" for future research:

1. Heat transfer enhancement via LVGs is a passive technique. Due to the continued development of moving mesh technique, semi-passive heat transfer enhancement techniques deserve more attentions from the researches. The LVGs will be subject to forced vibration due to the fluid pressure, which will further increase the disturbance of the flow field. The disturbance due to vibrations of LVGs can be combined with the swirling secondary flow and jet impingement to further enhance heat transfer.
2. If the LVGs on the fins are manufactured by punching, they will affect the fin efficient to

some extent [46]. In order to overcome this drawback, Hirokazu [46] cut the LVG to two or three sections and studied its performance. The results showed that the fin efficiency is slightly increased and the heat transfer enhancement is increased by 5% and 8% for two and three sections, respectively. Meanwhile, the pressure drop is also slightly decreased. However, the process of manufacturing fins with LVGs becomes complicated when the LVGs are cut into more sections.

3. Further observation of the flow field near the LVGs indicated that when the LVG is cut into two sections, the first section generates the longitudinal vortex while the second section accelerates the fluid and guides it to the tube in the next row. Based on the above analysis, the LVGs can be improved from the following two aspects. The first aspect is to cut the LVGs to two sections so that they are easy to manufacture and heat transfer can be further enhanced. The second aspect is to combine the delta and rectangular winglets, instead of using only one of them. The first half of the LVG is responsible to generate the longitudinal vortex so that the delta winglet is more appropriate. The second half of the LVG is to guide the flow and accelerate so that rectangular winglet works better. It can be expected that the heat transfer in the fin-and-tube heat exchanger will be further enhanced and the pressure drop can be decreased by using the composite winglets. The heat exchangers with high efficiency and low pressure drop can be obtained.

4. The placement (CFU and CFD) and the locations of the LVGs need to be optimized. Our preliminary study indicated that under the same number of LVGs, the pressure drop can be decreased by 8~10% by changing the placement and locations of the LVGs while keep the same level of heat transfer enhancement. For different heat exchangers, the optimized placements and locations differ. Therefore, it is necessary to thoroughly study the placements and locations of LVGs on various heat exchanger structures so that high-efficiency heat transfer enhancement structures can be obtained.

**Acknowledgements**

The present work is supported by the National Natural Science Foundation of China (grant numbers: U0934005 and 51176155), National Basic Research Program of China (973 Program) under grant number 2011CB710702 and the U.S. National Science Foundation (grant number: CBET-1066917). YLH also thanks her former students, Drs. Y. G. Lei, P. Chu and L. T. Tian for their fruitful contributions on this subject during their doctorial studies.